\documentclass[twocolumn]{aastex631}
\usepackage{CJKutf8} 
\usepackage{amsmath}
\usepackage{comment}
\usepackage{physics}
\usepackage{textcomp}

\newcommand\Msol{M$_{\odot}$}

\usepackage{soul}
\usepackage{threeparttable}
\received{March 02, 2024}
\revised{May 16, 2024}
\accepted{May 20, 2024}

\submitjournal{AJ}

\shortauthors{Khim et al.}
\shorttitle{NSCs in LSB Galaxies}

\begin{document}

\title{Properties of Nuclear Star Clusters in Low Surface Brightness Galaxies}
  
\correspondingauthor{Donghyeon J. Khim}
\email{galaxydiver@arizona.edu}

\author[0000-0002-7013-4392]{Donghyeon J. Khim}
\affiliation{Steward Observatory and Department of Astronomy, University of Arizona, 933 N. Cherry Ave., Tucson, AZ 85721, USA}

\author[0000-0002-5177-727X]{Dennis Zaritsky}
\affiliation{Steward Observatory and Department of Astronomy, University of Arizona, 933 N. Cherry Ave., Tucson, AZ 85721, USA}

\author[0000-0002-2527-8899]{Mika Lambert}
\affiliation{Department of Astronomy \& Astrophysics, University of California, Santa Cruz, 1156 High Street, Santa Cruz, CA 95064, USA}

\author[0000-0001-7618-8212]{Richard Donnerstein}
\affiliation{Steward Observatory and Department of Astronomy, University of Arizona, 933 N. Cherry Ave., Tucson, AZ 85721, USA}

\begin{abstract}
Using the SMUDGes and SDSS catalogs, and our own reprocessing of the Legacy Surveys imaging, we investigate the properties of nuclear star clusters (NSCs) in galaxies having central surface brightnesses as low as 27 mag arcsec$^{-2}$. We identify 273 (123 with known redshift) and 32 NSC-bearing galaxies in the two samples, respectively, where we require candidate NSCs to have a separation of less than 0.10$r_e$ from the galaxy center. We find that galaxies with low central surface brightness ($\mu_{0,g} > 24$ mag arcsec$^{-2}$) are more likely to contain an NSC if 1) they have a higher stellar mass, 2) a higher stellar to total mass ratio, 3) a brighter central surface brightness, 4) a larger axis ratio, or 5) lie in a denser environment. Because of the correlations among these various quantities, it is likely that only one or two are true physical drivers. 
We also find scaling relations for the NSC mass with stellar mass ($M_{NSC}/$\Msol$ = 10^{6.02\pm0.03}(M_{*,gal}/10^{8} $\Msol$)^{0.77\pm0.04}$) and halo mass ($M_{NSC}/$\Msol$ = 10^{6.11\pm0.05}(M_{h,gal}/10^{10} $\Msol$)^{0.92\pm0.05}$), although it is the scaling with halo mass that is consistent with a direct proportionality. In galaxies with an NSC, $M_{NSC} \approx 10^{-4}M_{h,gal}$. This proportionality echoes the finding of a direct proportionality between the mass (or number) of globular clusters (GCs) in galaxies and the galaxy's total mass. These findings favor a related origin for GCs and NSCs. 
\end{abstract}


\section{Introduction}
\label{sec:intro}

Extremely concentrated stellar populations, referred to as nuclear star clusters (NSCs) or stellar nuclei, exist at the center of galaxies of diverse morphology \citep[e.g.,][]{Lauer2005, Georgiev2014} and environment \citep[e.g.,][]{Baldassare2014, sanchez2019b} that range from local group spheroidals \citep{Crnojevic2016} to those that are significantly more massive than the Milky Way \citep[$M_{*,gal} \sim 10^{11.8}$ \Msol;][]{Neumayer2012}. How host galaxy properties affect the likelihood and properties of an NSC and how such dependencies constrain NSC formation models remain open questions and are our focus.

NSCs are not found in all galaxies even when deep images are examined \citep[e.g.,][]{Lim2018}. Determining whether NSC formation is genuinely stochastic or depends on galaxy properties is of particular interest in assessing formation scenarios.
Already, the fraction of galaxies that host an NSC, the occupation fraction, has been shown to depend on stellar mass \citep{denBrok2014, Ordenes2018, neumayer}, galaxy shape \citep{Lisker07}, central surface brightness \citep{lim}, morphology \citep{ 2020MNRAS.491.1901H}, and environment \citep{sanchez2019, denBrok2014, Lim2018, neumayer}. 
Some of these relationships are complex, such as the one with stellar mass, which peaks at a stellar mass ($M_{*,gal}$) of $\sim 10^{9}$\Msol\ \citep{Cote2006, sanchez2019}. Others are more controversial, such as those involving morphology and environment, where conflicting findings exist (\cite{neumayer} for morphology and \cite{Baldassare2014, georgiev, Georgiev2014} for environment).
 
As with the occupation fraction, studies have found that the mass of the resulting NSC, $M_{NSC}$, is related to host galaxy properties. In this case, however, the only identified relation to date is that with the host galaxy stellar mass  \citep{neumayer}. 
Interestingly, that relation is sublinear, where $M_{NSC}$ rises more slowly than $M_{*,gal}$, suggesting that the two stellar populations formed via different channels. This behavior is reminiscent of that seen for the total mass of globular clusters (GCs) in a galaxy, where the GC mass tracks the total mass of the host galaxy more closely than it tracks the stellar mass \citep{blakeslee, Spitler2009, Harris2017}. The literature has not established whether this behavior is fully mimicked by NSCs, but the possible connection to GCs is a critical one for NSC formation models.

There are two broad families of NSC formation scenarios.
The first suggests that NSCs are created through the infall and merging of GCs \citep[e.g.,][]{tremaine1975, lotz, CDMB2009, gnedin}. Implications of this scenario include predictions regarding the properties of the initial and surviving GCs. For example, the observed lack of massive GCs in the inner region of galaxies \citep[e.g.,][]{lotz, CDMB2009} has been attributed to the effect of dynamical friction, which is required to cause the inward spiraling of GCs that would merge to create an NSC. The second suggests that NSCs were formed by extreme in-situ star formation \citep[e.g.,][]{Bailey1980, mihos94, bekki01, Seth2006, Walcher2006}, perhaps with infalling gas from a gas-rich merger leading to highly centralized star formation. Implications of this scenario might include that NSCs have a wide range of ages, reflected by their colors, and show weak, if any, correlations with host galaxy properties.

The correct formation scenario will reproduce whatever trends are identified and confidently confirmed between the occupation fraction, $M_{NSC}$, and host galaxy properties. 
We seek to further establish, or refute, the trends we have already cited, and identify any remaining undiscovered trends. Both of these goals are made easier by exploring the widest range of galaxy properties possible. With recent advances, we are able to significantly extend the study of NSCs to host galaxies with extremely low central surface brightness. Particularly interesting in this regard are ultra-diffuse galaxies \citep[UDGs; $\mu_{0,g} \ge 24$ mag arcsec$^{-2}$, $r_e > 1.5$ kpc;][]{2015vanDokkum} because they are outliers in various galaxy relationships \citep{Beasley2016, Li2023}. As such, they may help break degeneracies among relationships and clarify which of these are more fundamental than others. 

In our preceding study \citep{lambert}, we identified NSCs in the Systematically Measuring Ultra-Diffuse Galaxies (SMUDGes) set of UDG candidates \citep{smudges, smudges2, smudges3, smudges5}. Here we conduct a more comprehensive analysis of the scaling relations between NSC mass, stellar mass, and dark matter halo mass. 
We also extend our investigation to explore the dependencies of the NSC occupation fraction, requiring an expansion of the surface brightness range. We broaden our sample by incorporating galaxies from the Sloan Digital Sky Survey (SDSS), which are typically of higher surface brightness and mass. We develop a pipeline for fitting galaxies with single S\'ersic profiles \citep{sersic} and simultaneous decomposition into multiple components to consistently and quantitatively identify NSCs. Then, we proceed to establish the connections between NSCs and their host galaxies in terms of the NSC occupation fraction and the scaling relations between $M_{NSC}$ and host properties. Where possible, we compare our results with the previous known relations to provide confidence in the identified relationships. We investigate two primary questions: are the host galaxies of NSCs distinct from the broader sample of galaxies and do the characteristics of NSCs vary based on the properties of their host galaxies.

This paper is organized as follows. In \S \ref{sec:method}, we describe our methodology, including how we identify NSCs, measure the stellar masses for the hosts and NSCs, estimate the halo masses of the hosts, and parameterize their environments. In \S \ref{sec:results}, we examine and present the relationships between the properties of host galaxies and NSCs. In \S \ref{sec:discussion}, we discuss our findings in the context of NSC formation scenarios. Finally, we summarize our findings in \S \ref{sec:summary}. Through the paper, we utilize a standard WMAP9 cosmology \citep{wmap9}, although our findings are insensitive to the current range of uncertainty in cosmological parameters. Magnitudes are measured from the Legacy Survey data and are in the AB system \citep{oke1,oke2}.

\section{Methodology}
\label{sec:method}

\subsection{The data}
\label{sec:data}
Our sample consists of data from two separate surveys that we analyze in a consistent manner. First, we have targets from the Systematically Measuring Ultra-Diffuse Galaxies (SMUDGes) complete catalog \citep{smudges5}, whose NSC population was initially presented by \cite{lambert}. 
Second, we have targets from the SDSS catalog DR16 \citep{2020ApJS..249....3A, 2017AJ....154...28B},
aiming both to expand the range of host galaxy central surface brightnesses and increase the number of galaxies with NSCs for which we have measured distances. 

SMUDGes were originally selected to have low central $g$-band surface brightness, $\mu_{0,g}$ $\geq$ 24 mag arcsec$^{-2}$, and large effective angular size, r$_e$ $\geq$ 5.3 arcsec. The size criterion corresponds to $r_e \geq 2.5$ kpc at the distance of the Coma cluster\footnote{We adopt an angular diameter distance of 98 Mpc for the Coma cluster to maintain consistency with \cite{vdk15a}.}, significantly larger than the typical 1.5 kpc minimum size definition for UDGs, although many candidates turn out to be nearer and therefore physically smaller. After applying a set of criteria, including a final visual examination, that sample consists of 6,805 UDG candidates \citep{smudges5}. 

\cite{lambert} applied a few additional selection criteria to refine that sample for their NSC analysis: 1) a stricter color criterion (0 $< g-r <0.8$) to eliminate potential background interlopers and candidates with an unphysical blue color, and 2) an additional angular size criterion ($r_e < 26$ arcsec) to exclude nearby galaxies that are less likely to be UDGs. 
Their final sample size drawn from SMUDGes is 6,542 galaxies. 
Within this dataset, we have spectroscopic redshifts for 226 galaxies from a variety of published sources \cite[see][]{kadowaki21} and our ongoing work both at optical (Ascencio et al. in prep.) and radio (Karunakaran et al. in prep.) wavelengths. Additionally, for 1,307 of these galaxies, we have estimated redshifts based on their likely association with galaxy clusters or groups \citep{smudges5}. In cases where a galaxy has both a spectroscopic and an estimated redshift, we adopt the spectroscopic value.

Our SDSS targets are those designated\footnote{See \url{https://live-sdss4org-dr16.pantheonsite.io/spectro/catalogs/} 
for a detailed description of this category, which aims to identify objects with secure spectroscopic redshifts.} as `sciencePrimary',  
and have a redshift uncertainty $<$ 0.001, no redshift warning flag (`zwarning'), and a median signal-to-noise ratio (S/N or SNR) $>$ 2. The SNR is measured per pixel across the full spectrum.
Furthermore, we require that the recessional velocity be greater than 1,000 km s$^{-1}$, to select objects that are outside the Milky Way, and 
that the size of the point spread function (PSF), as defined by the seeing in the particular image, corresponds to $<$ 500 pc, to ensure that we resolve any nuclear structure (bulge) that has an effective radius $>$ 500 pc. Our choice to resolve central structures larger than 500 pc is set roughly by the effective radius of the smallest galaxy bulges \citep{gadotti}. 
With the resolution criteria, the redshift range of the SDSS sample extends up to 0.023 (cz $\sim$ 7,800 km s$^{-1}$), but we do not apply a defined upper redshift cut.

The SDSS catalog does provide estimates of the color and angular size of these objects obtained from their own images. However, given the Legacy Survey's superior spatial resolution (0.262 arcsec pixel$^{-1}$ vs. 0.396 arcsec pixel$^{-1}$) and greater sensitivity \citep{dey}, we perform our own measurements of these parameters, as described in \S \ref{sec:fitting}, and then
apply the color and angular size cuts (0 $< g-r <0.8$ and $r_e < 26$ arcsec) that we applied to the SMUDGes sample.
Additionally, for targets that are not best modeled as isolated point sources, we apply an additional cut requiring $r_e > 3$ arcsec,
which is roughly twice the
median value of the FWHM (full-width half maximum) of the PSF, to remove galaxies that are too small on the sky for us to accurately resolve multiple components.

\subsection{Image Extraction and Masking}

We extract working images from the Legacy database and mask contaminants as summarized here and described in more detail by \cite{lambert}. We extract 200$\times$200 pixel (52.4 $\times$ 52.4 arcsec) $r$- and $g$-band images of each selected target from the 9th data release (DR9) of the Legacy Survey \citep{dey}. This selection ensures that the obtained cutouts encompass an ample field of view, allowing for adequate background estimation.
To generate image masks, we utilize the Source Extractor Python library \citep[SEP, see][for details]{sep} based on Source Extractor \citep{bertin}. First, we use SEP to subtract the spatially varying background and measure the background noise. We use a 64$\times$64 pixel mesh (16.8 $\times$ 16.8 arcsec) to measure the background. We use this background model only when identifying objects to mask. It will not affect the photometric modeling of our galaxies.

We then employ SEP again to identify objects, which are defined as groups of at least 5 adjacent pixels where each has a flux that is $\ge 1.5\sigma$ above the background. We mask these objects, except for the target galaxy itself. If the masked regions cover more than 50\% of the entire image area, we exclude the target from further analysis and discussion. At this stage, our sample consists of 6,538 SMUDGes and 4,746 SDSS targets.

\begin{figure*}[ht]
    \centering
    \includegraphics[scale=0.42]{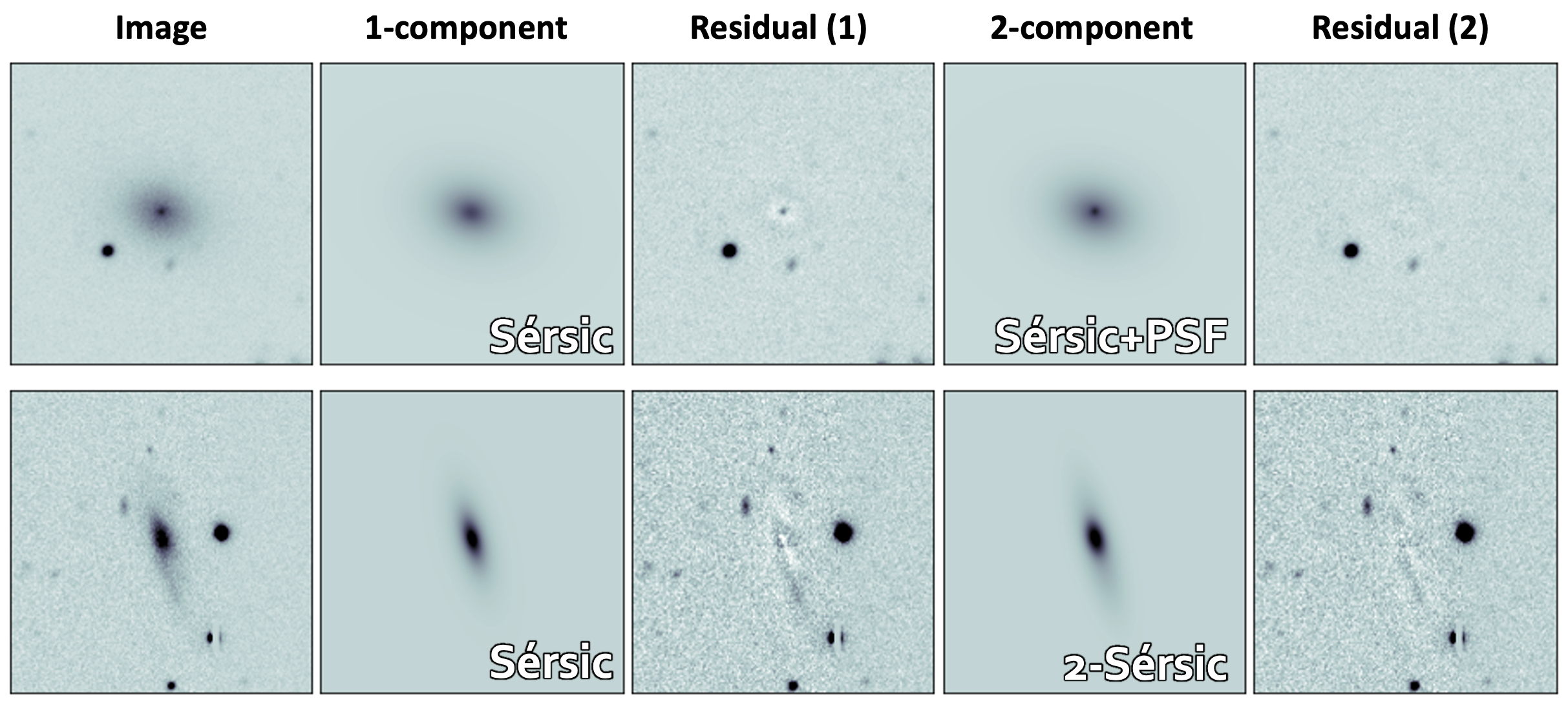}
    \caption{Model and residual images for two SDSS galaxies. In the upper row, we show a set of images corresponding to the first galaxy in our sample (sorted by SDSS `SpecObjID'), for which the S$_1$ + N$_{1,PSF}$ model is statistically preferred. In the lower row, we show the corresponding set of images for the very first galaxy in our SDSS catalog that is classified as double S\'ersic. In each row, the images are from left to right: the original $r$-band image, the best-fit single S\'ersic model, the difference between the $r$-band image and best-fit single S\'ersic model, the best-fit two-component model (S\'ersic + PSF model for the upper row, double-S\'ersic for the bottom row), and the difference between the $r$-band image and best-fit two-component model.
    The display scale of the images in the third and fifth columns is stretched by a factor of 2 to highlight the residuals. The NSC in the upper row is evident in the residuals from the single component fit, and well modeled by the two component one. The central structure is somewhat ambiguous for the second galaxy, but there is no clear evidence for an NSC. }
    \label{fig:models}
\end{figure*}

\begin{deluxetable}{lrr}
\caption{Modeling Summary}
\label{tab:models}
\tablehead{\colhead{1st component}&\colhead{2nd component}&\colhead{3rd component}}
\startdata
\multicolumn{3}{c}{\underbar{GALFIT Stage 1}} \\
\\
S$_1$ & ... & ... \\
\\
\multicolumn{3}{c}{\underbar{GALFIT Stage 2}} \\
\\
S$_1$ & ... & ... \\
S$_1$ & S$_2$ & ... \\
S$_1$ & N$_{1,PSF}$ & ... \\
S$_1$ & S$_2$ & N$_{1,PSF}$\\
S$_1$ & N$_{1,PSF}$ & N$_{2,PSF}$ \\
N$_{1,PSF}$ & ... & ... \\
\\
\enddata
\tablecomments{The two GALFIT fitting stages and the models used in each (see \S \ref{sec:fitting}). The initial parameters used for fitting in the second stage are those of the best-fit model from the first stage. S stands for S\'ersic models and N for unresolved source, potentially NSCs.
}
\end{deluxetable}

\subsection{Model fitting}
\label{sec:fitting}
We employ the photometric model fitting software package GALFIT \citep{Peng_2010} to separate the structural components and to measure their photometric parameters. We adopt the PSF supplied by the Legacy Survey and calculate the pixel-by-pixel uncertainties ($\sigma$-images) using the Legacy Survey provided inverse-variance images. We adopt a flat background to avoid oversubtracting the wings of the galaxy \citep[a well-known problem in galaxy photometry, e.g.,][]{moustakas}. We set the convolution box size for GALFIT to half the length of the image.

The GALFIT fitting results are often significantly influenced by the choice of initial parameters, principally due to the diffuse and faint nature of our targets. This sensitivity is compounded by the possible existence of a central component, such as a bulge or NSC. 
To address this difficulty, we implement a two-stage multi-component fitting procedure,
similar to \cite{lambert} and summarized below. Additionally, at each fitting stage, we perform multiple fits, adjusting the initial parameter values to enhance the robustness of our analysis. 
Figure \ref{fig:models} displays examples of models and residual images created by GALFIT for single S\'ersic models and two-component models for two SDSS galaxies, serving as a brief introduction to the motivation and the results of our GALFIT pipeline. The single S\'ersic models (the second column) produce inadequate residual maps in the central region (the third column), even beyond those due to an NSC or a bulge in these two cases. Two-component models (fourth column) yield improved residuals in the central region in both (fifth column). Further details regarding these fitting stages will be provided in the following sections.
A summary of these steps is described in Table \ref{tab:models}, with additional details provided subsequently.

\subsubsection{The first stage: a single S\'ersic component}

In the first stage, we fit a single S\'ersic profile to each galaxy, avoiding nearby or overlapping objects or components such as an NSC or bulge. 
To achieve this, we augment the mask to include any central region containing at least 5 adjacent pixels that have a corresponding surface brightness 1.5 times (0.44 mag) brighter than 24 mag arcsec$^{-2}$. 

We provide GALFIT with the target image, augmented image mask, PSF model, $\sigma$-image, and a constraint file that sets the search range for each of the free parameters. The free parameters are the following: central position, S\'ersic index (n), effective radius ($r_e$), magnitude (m), axis ratio (AR or b/a), position angle (PA), and the background level. 
We constrain the center of the S\'ersic components to lie within a 40 by 40 pixel square centered on the image. Because our sample mainly consists of low luminosity galaxies, we set the upper limit of the S\'ersic index to be 2.0. SMUDGes have $\langle n \rangle < 1$ \citep{smudges3} and small spheroidals tend to also have low $n$ \citep{caon}. Furthermore, this choice is empirically validated by the modest number of targets for which this upper bound on $n$ turns out to be the best-fit value (131 or 2\% among SMUDGes targets, and 529 or 11\% among SDSS targets). We also set a lower limit on the axis ratio of the S\'ersic component of 0.3 to prevent unrealistically elongated models. Note that the SMUDGes sample has an axis ratio threshold of $0.34 < $ AR \citep{smudges5}.

To mitigate the impact of our initial parameter choices, we conduct the fitting process six times using various initial guesses. We use combinations of two different effective radii (30 and 50 pixels) and three surface brightness values at $r_e$ (25, 28, and 31 mag arcsec$^{-2}$). 
We compute the reduced chi-squared statistic, $\chi^2_\nu$, within a circular region of radius 50 pixels ($\sim 13.1$ arcsec) centered on the image center and select the model with the smallest $\chi^2_\nu$ value. GALFIT occasionally produces a model fit with huge final parameter uncertainties that has a statistically acceptable (i.e., confidence level of more than 90\%) $\chi^2_\nu$ value. We only consider models where $r_e > 2\sigma_{r_e}$ as meaningful.
The results obtained from this fitting process are used as input for the next GALFIT stage, which involves more complex models.

\subsubsection{The second stage: multiple components}

In the second stage, we simultaneously decompose the galaxy image into multiple components. We employ six independent model classes to explore the potential presence of a secondary component, such as a bulge or NSC (See Table \ref{tab:models}). These models are motivated by the following scenarios: 
1) a single S\'ersic profile (a galaxy with no unresolved nuclear source); 
2) a pair of S\'ersic profiles (a galaxy with an additional extended central source, such as a separate bulge component);
3) a S\'ersic profile with a PSF profile (a galaxy with an unresolved central source, such as an NSC); 
4) a pair of S\'ersic profiles with a PSF profile (a galaxy with an extended central source, such as a separate bulge component, and an unresolved central source);
5) a S\'ersic profile with two PSF profiles (a galaxy with two unresolved central sources); and
6) a single PSF profile (``dark'' galaxies that have no detectable S\'ersic component, only an unresolved cored source). 
We add the sixth model to those employed by \cite{lambert} because some of the SDSS targets 
within our redshift selection range (i.e., naively beyond the Milky Way) appear unresolved. 
We unmask the center of the host galaxy and all other sources within $0.5r_e$ for this second fitting stage.

\begin{deluxetable*}{lc|rrrrc|rr}  
\caption{Classification Summary}
\label{tab:class}
\tablehead{ & & \multicolumn{4}{c}{SMUDGes} & & \multicolumn{2}{c}{SDSS} \\
\colhead{Class} & & \colhead{N} & \colhead{$r_p/r_e < 0.10$} & \colhead{N$_{Dist}$} & \colhead{$r_p/r_e < 0.10$}
& & \colhead{N$_{Dist}$} & \colhead{$r_p/r_e < 0.10$} }
\startdata
\underline{No unresolved} & & & & & & \\
\underline{source} & & & & & & \\
 & & & & & & \\
S$_1$ & & 2700 & ... & 649 & ... &   & 16 & ... \\
S$_1$+S$_2$ & & 1558 & ... & 283 & ... & & 627 & ... \\
S$_1$+N$_{1,PSF} ?$ & & 1338 & ... & 312 & ... & & 186 & ... \\
Total & & 5596 & ... & 1244 & ... & & 829 & ...  \\
 & & & & & & \\
\underline{Unresolved} & & & & & & \\
\underline{source} & & & & & & \\
 & & & & & & \\
S$_1$+N$_{1,PSF}$ & & 451 & 202 & 160 & 91 & & 30 & 24 \\
S$_1$+N$_{1,PSF}$ & & 127 & 71 & 51 & 32 & & 12 & 8 \\
\ \ \ \ + N$_{2,PSF}$ & & & & & & \\
Total & & 578 & 273 & 211 & 123 & & 42 & 32 \\
 & & & & & & \\
\underline{Bare PSF} & & 0 & ... & 0 & ... & & 18 & ... \\
 & & & & & & \\
\underline{Failed fitting} & & 364 & ... & 72 & ... & & 3869 & ... \\
 & & & & & & \\
Total & & 6538 & 273 & 1527 & 123 & & 4746 & 32 \\
\enddata
\tablecomments{Results for SMUDGes and SDSS candidates satisfying the additional selection criteria described in the text. For SMUDGes candidates, we present results both for the full sample and the subsample with estimated distances. All SDSS galaxies have spectroscopic redshifts and therefore estimated distances. We present the numbers of systems where the 
unresolved source lies within a normalized projected separation, $r_p/r_e$, from the S$_1$ component that is $< 0.10$, where $r_e$ is measured in the initial fitting of the single S\'ersic model. These are the galaxies that we consider to contain an NSC.
}
\end{deluxetable*}

We refer to the component used to model the underlying galaxy as S$_1$, which is represented with a S\'ersic profile. In the second and fourth models, we refer to the second, more compact S\'ersic component aimed at representing a resolved central excess as S$_2$.
For models with an unresolved central source (i.e., PSF), we denote this component as N$_{1,PSF}$. For the fifth model category, we denote the unresolved component situated nearer to the center of the S$_1$ component as N$_{1,PSF}$ and the more distant as N$_{2,PSF}$.

In this GALFIT stage, the free parameters are the central positions, S\'ersic indices (n), effective radii ($r_e$), magnitudes (m), axis ratios (AR), and position angles (PA) for S$_1$ and S$_2$; the positions and amplitudes for N$_{1,PSF}$ and N$_{2,PSF}$; and the background level. 
As in the first stage, we set the upper limit of the S\'ersic index for S$_1$ to be 2.0. On the other hand, the S\'ersic index of the compact S\'ersic component (S$_2$) is allowed to range up to 5.0, considering its potential morphological resemblance to a classical bulge \citep[$n=4$, a de Vaucouleurs profile introduced by][]{dev}. Furthermore, we require that the effective radius of each S\'ersic component exceed 0.75 times the size of the PSF to establish a clear distinction between what we categorize as a resolved component and an unresolved one.
Finally, the constraints on positions and axis ratios remain consistent with those employed in the first GALFIT stage.

As in the first fitting stage, we conduct multiple fits for each model class using different initial parameters. 
For S$_1$, we utilize the parameters from the best-fit model in the first stage, except for the central surface brightness. We utilize the initial guess for the brightness ranging from three magnitudes fainter to brighter.
%Regarding the initial guess for the brightness of each component, we consider permutations of the previous fit and values that are three magnitudes fainter and brighter. 
While the initial position for S$_1$ is determined based on the best-fit model in the first stage, we assign two initial positions for the rest of the components. The initial guesses are either the image center or the brightest pixel within the position constraints, defined as a 40 by 40 pixel square centered on the image.
We set the initial guess of the size of S$_2$ as five pixels. We find no improvement in fitting when varying $r_e$ for S$_1$, so we simply use the result from the initial model fit as the initial guess. 
Because S$_2$ is a more compact component than S$_1$, the initial S\'ersic index is set to 3.

Before comparing the resulting models, we exclude models that are rejected with at least 90\% confidence given their $\chi^2_\nu$. 
Our primary focus is on the central region of the galaxy, and as such, we assess $\chi^2_\nu$ in a circular area of radius  $0.5r_e$ centered on the galaxy. The values for $r_e$ and the central position needed to evaluate $\chi^2_\nu$ are derived from the single S\'ersic model fit obtained in the first fitting stage.
Once again, we only consider models with $r_e > 2\sigma_{r_e}$ to be meaningful. If no models meet both the $\chi^2_\nu$ and $r_e > 2\sigma_{r_e}$ criteria, then the galaxy is considered to have failed our fitting procedure, indicating that we cannot confidently describe it with the available combination of S\'ersic and PSF profiles.

We implement all of this on the $r$-band data. For the $g$-band model fits we leave free only the magnitudes of each component, using the $r$-band results for the structural parameters. Where the $g$-band model fails to provide a good fit or when the magnitude uncertainty $>$ 0.2 mag, we reject the model.  
Consequently, all our models have significant detections in both $r$- and $g$-bands.
Where we quote photometric properties of the galaxies, such as $\mu_0$ or $g-r$, these refer to the measures for the S$_1$ component and exclude any NSC or other components that are detected.
We correct all of our flux measurements for dust extinction based on the SFD \citep{SFD} dust maps using the \texttt{dustmaps} package \citep{green}\footnote{\url{https://dustmaps.readthedocs.io/en/latest/index.html}.}.

\subsection{Classification}
\label{sec:classification}

While $\chi^2_\nu$ values offer a measure of the goodness of fit, they are not suitable for comparing models with different degrees of freedom. A model with more fitting parameters will naturally achieve a better fit to the given data. It is essential to consider this added flexibility when evaluating whether there is statistical support for the more intricate model. The Akaike Information Criterion (AIC), as introduced by \cite{AIC}, is one such formulation that introduces a penalty for models with increased complexity.

We implement the small-sample corrected AIC, known as the AICc criterion \citep{sugiura}. The AICc is represented by the equation:
\begin{equation}
AICc= \chi^2+2p+ \frac{2p(p+1)}{N-p-1}
\end{equation}
where $p$ denotes the quantity of model parameters, and $N$ is the number of fitted data points.

The model exhibiting a lower AICc value is the preferred one statistically, and the larger the difference ($\Delta$AICc), the greater the confidence in distinguishing between them. As AICc values represent likelihoods and follow a distribution similar to $\chi^2$, it is possible to determine the confidence level associated with any specific $\Delta$AICc value. In our context, a 3$\sigma$ confidence level corresponds to $\Delta$AICc $= 11.83$. We adopt this threshold to assess if the best-fitting two-component models (S$_1 + $S$_2$ and S$_1 +$N$_{1,PSF}$) are significantly preferred over the best-fitting one-component model (S$_1$), as also implemented by \cite{lambert}. Following the same logic, three-component models must exhibit statistical preference over two-component models with a confidence level surpassing $3\sigma$.

Eighteen SDSS targets are best fit with a single PSF and no underlying S\'ersic profile, despite being presumed to be outside the Milky Way and therefore not stars (recessional velocity $>$ 1,000 km s$^{-1}$). Because these objects do not have a clear S$_1$ component, we cannot calculate AICc within 0.5$r_e$ for our comparison of models. Instead, we compare AICc values within a circular region of radius 20 pixels ($\sim 5.2$ arcsec) centered on the image center and select the model with the smallest AICc values. We will return to these objects in \S \ref{sec:bare_psf}.

For targets with best-fitting two-component models, we employ the same $\Delta$AICc confidence threshold to distinguish between the two-component models. We classify the target as S$_1$ + N$_{1,PSF}$ if that model has a lower (better) AICc value with $\Delta$AICc $\geq 11.83$ compared to the S$_1$ + S$_2$ model. We classify the target as S$_1$ + N$_{1,PSF}?$ if the target prefers the S$_1$ + N$_{1,PSF}$ model, but the AICc difference is not significant ($\Delta$AICc $< 11.83$) compared to the S$_1$ + S$_2$ model. We apply the same method to distinguish among possibilities for targets favoring the best-fitting three-component models.

We find that models that include an S$_2$ component with a large S\'ersic index, $n \ge 4$, and a small size, $r_e < 10$ pixel, are usually visually indistinguishable, and not statistically different, from those with N$_{1,PSF}$. 
Therefore, we reclassify cases where the two-component models are statistically preferred over a one-component model at a confidence level exceeding 3$\sigma$ and the S$_2$ component satisfies the conditions $n\ge4$, $r_e < 10$ pix (2.62 arcsec), and lies within 5 pixels of the competing N$_{1,PSF}$ component, 
as S$_1$ + N$_{1,PSF}$.
This is consistent with the treatment adopted by \cite{lambert}.

To refine the sample to the most robust and best measured nuclear sources, we implement additional selection criteria for galaxies containing unresolved sources (S$_1$ + N$_{1,PSF}$ and S$_1$ + N$_{1,PSF}$ + N$_{2,PSF}$): 1) the color of the unresolved source must lie in the range 0 $< g-r < 0.8$, 2) the photometric uncertainty of 
the color of the unresolved source must be below 0.2 mag, and 3) the S\'ersic $n$ of S$_1$ must be $<2$. The third criterion is rarely invoked (occurring only 6 and 1 times in total for SMUDGes and SDSS, respectively) but is implemented to identify galaxies for which the derived photometric parameters may be poorly measured.
We reclassify galaxies that fail these criteria as S$_1$ + N$_{1,PSF}?$.

We apply the same classification scheme to both the SMUDGes and SDSS samples, and present the summary of our classifications in Table \ref{tab:class}. For the SMUDGes sample, we also list the numbers of targets with measured or estimated distances (N$_{Dist}$). All SDSS galaxies have spectroscopic redshifts and therefore measured distances.

\cite{lambert} applied a radial selection cut, $r_p/r_e < 0.1$, to help distinguish true NSCs from contamination. Here, $r_p$ represents the projected radial offsets between S$_1$ and N$_{1,PSF}$, and $r_e$ denotes the effective radius of the host galaxy measured during our initial fitting pass. 
This strict radial selection cut ensures a modest contamination fraction (0.15) for the NSC sample. We apply the same radial selection cut to our samples. We provide the number of galaxies having unresolved sources both with and without this additional selection cut in Table \ref{tab:class}. Henceforth, we refer to the subset of unresolved sources that satisfy $r_p/r_e < 0.1$ as NSCs and present results using only these sources. 
We find 273 (123 with known redshift) and 32 NSC-bearing galaxies in the SMUDGes and SDSS samples, respectively. In the SMUDGes sample, we find 26 NSC-bearing galaxies that satisfy the conventional UDG criteria ($r_e \geq$ 1.5 kpc, $\mu_{0,g} \geq$ 24 mag arcsec$^{-2}$). We find no UDGs in the SDSS sample because all of the SDSS galaxies in our sample have central surface brightness brighter than 24 mag arcsec$^{-2}$.

\subsection{Detection Limits}
\label{sec:completeness}

\begin{figure}
	\includegraphics[width=\columnwidth]{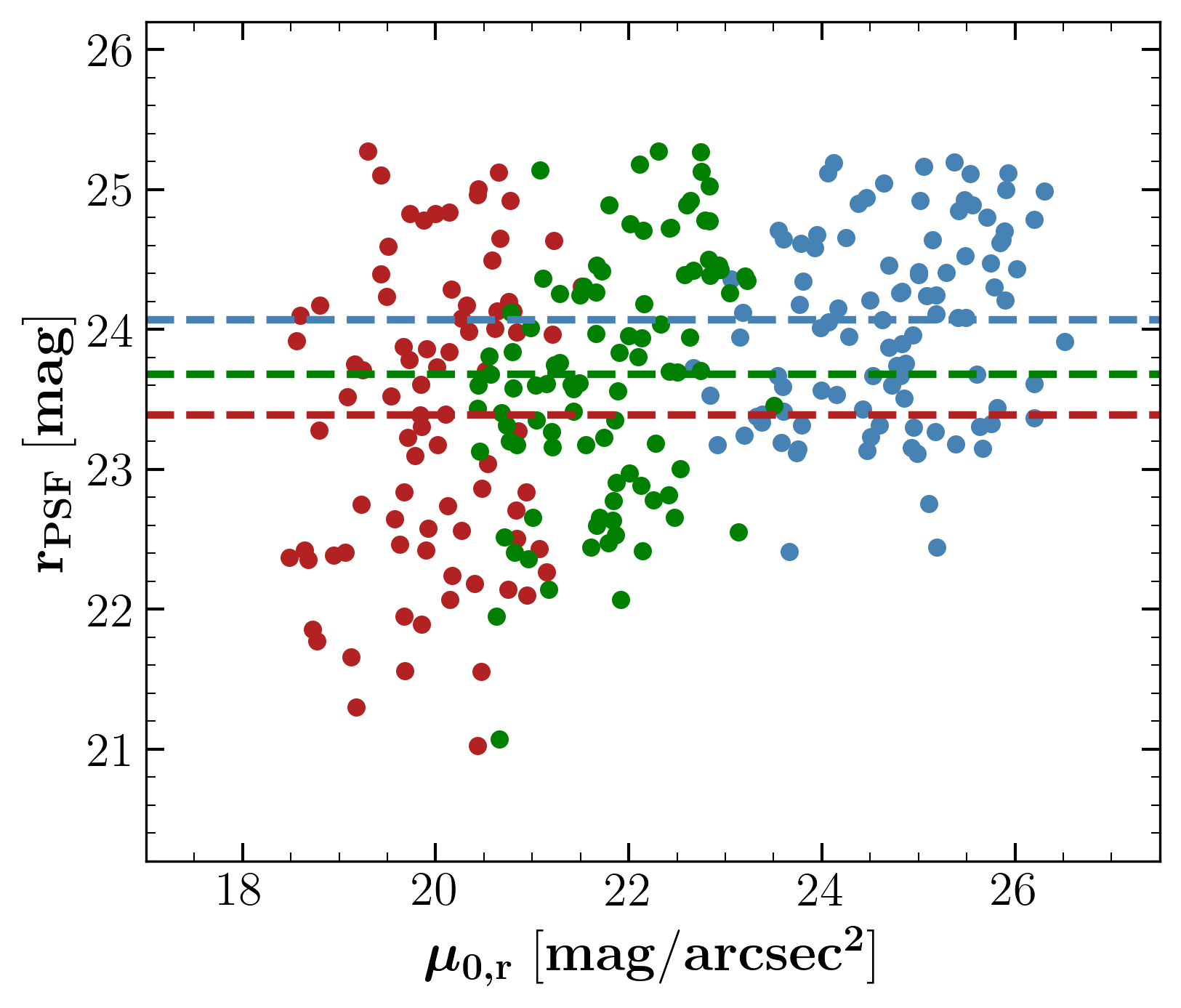}
    \caption{NSC detection limits in the $r-$band as a function of $r-$band central surface brightness, $\mu_{0,r}$, for three different sets of simulations. The analysis involves the placement and attempted recovery of a central point source in a set of single S\'ersic SMUDGes galaxies (blue), and versions of those galaxy images where the flux is increased by a factor of 16 (green) and 100 (red).
    Points represent the most luminous simulated point source that we failed to identify as an NSC in each galaxy. The horizontal lines mark the 50\% completeness magnitude for the three samples, ranging from 23.4 mag for the red points and 24.1 mag for the blue points.}
    \label{fig:detection_limit}
\end{figure}

Among the galaxies for which we find acceptable models but do not find an NSC, some may host an NSC that falls below our detection threshold. \cite{lambert} estimated the detection limit of their NSC classification by adding an artificial point source with varying brightness to a randomly selected set of 100 SMUDGes that are classified as single S\'ersic. The faintest artificial source that they detect with greater than 3$\sigma$ confidence ($\Delta$AICc $>$ 11.83) is defined to be the detection limit.
Their results are reprised in Figure \ref{fig:detection_limit} as the blue points and are appropriate for the SMUDGes targets presented here. 

Those results may not be appropriate for 
the SDSS galaxies, which have generally significantly higher central surface brightness ($\langle \mu_{0,r}\rangle = 21.2$ mag arcsec$^{-2}$) than do SMUDGes galaxies.
We expect a correlation between the detection limit and host central surface brightness because it becomes more challenging to detect an NSC against a brighter background. As such, we must extend the \cite{lambert} completeness results to higher central surface brightness. 

Unfortunately, we cannot simply redo the earlier test with the SDSS galaxies because there is a limited number of SDSS galaxies that are best-fit with a single S\'ersic profile. Instead, we use the same galaxies used by \cite{lambert} and amplify their flux by a factor of 16 to create one sample and by a factor of 100 to create another, corresponding to samples that are three and five magnitudes brighter, respectively. We then redo the \cite{lambert} completeness simulations.

We present the resulting detection limits for these higher surface brightness galaxies in Figure \ref{fig:detection_limit}. The three samples are distinguished by color in the Figure and serve to cover central surface brightnesses between 18 and 27 mag arcsec$^{-2}$,  extending beyond the range of our targets on either side. The 50\% completeness NSC magnitude drops by about one magnitude from the low to the high central surface ends of the explored range. 
The Figure highlights one reason why we might expect a larger number of undetected NSCs, and correspondingly a lower occupation fraction, for the SDSS sample. 

For the SMUDGes sample, we conclude that we are quite complete. We find no NSCs with $r > 24$ mag despite being sensitive to such NSCs in half the sample. As a test of this conclusion, we compare the slope of the $m_{nsc}-m_{gal}$ relation obtained from the full sample (0.784 $\pm$ 0.40) to that from a trimmed  ($m_{gal} < 19$) sample (0.777 $\pm$ 0.051), where we are even more likely to be complete because the NSCs are themselves brighter. The excellent agreement in the slopes demonstrates that we are not creating an artificial turnover due to incompleteness.

\begin{figure}
	\includegraphics[width=\columnwidth]{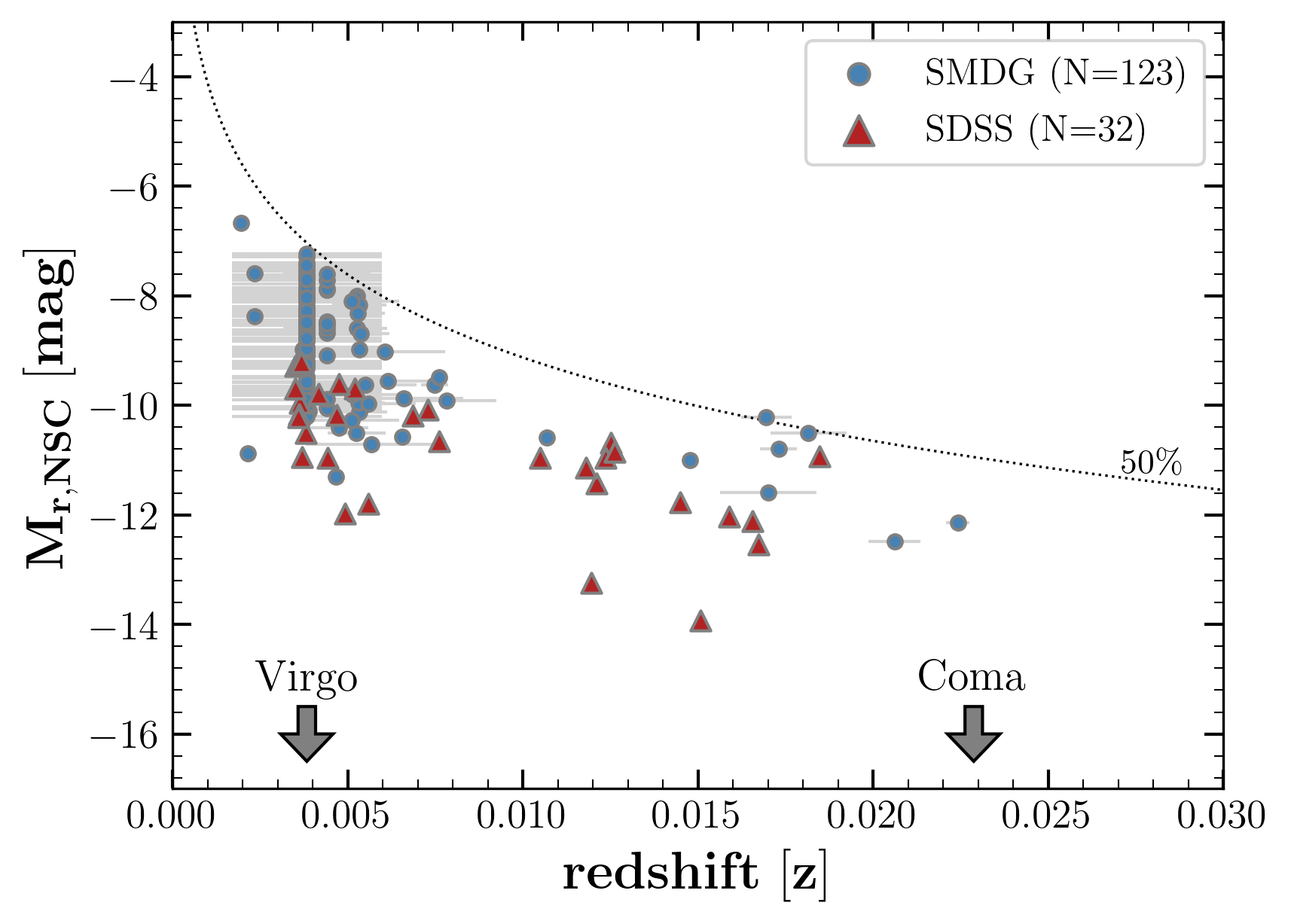}
    \caption{The absolute magnitude of NSCs in SMUDGes (blue) and SDSS (red) galaxies across redshift. The plotted curve represents the 50\% completeness threshold for the SMUDGes sample, which corresponds to the blue horizontal line illustrated in Figure \ref{fig:detection_limit}. For reference, we denote the redshifts corresponding to the Virgo and Coma clusters.}
    \label{fig:detection_limit_dist}
\end{figure}

A different way of assessing this issue is to examine the distribution among detected NSCs as a function of distance. 
Given our fixed apparent magnitude limit, we expect lower completion as distance increases. Indeed we see such behavior (Figure \ref{fig:detection_limit_dist}), but the limiting factor is not our sensitivity to NSCs. 
We reach this conclusion because only one NSC in the sample is fainter than our 50\% completeness limit. This result suggests that we are instead dominated by missing the fainter hosts, which would host these fainter NSCs, as a function of distance. 
This bias arises because SMUDGes selected systems using surface brightness, which is distance independent over this redshift range, and cutting on angular size. 
Given the rather universal S\'ersic profiles of these galaxies, the angular size cut effectively limits the host galaxy magnitude in a manner that depends on distance. We conclude that we are highly complete in NSC detection in the host galaxies we have, but are incomplete in the host sample. 
This type of incompleteness does not bias any of the results we present, but would bias estimates of the total numbers of NSCs in this volume derived from this sample.

\subsection{Stellar and Dark Matter Halo Mass Estimation}
\label{sec:dmh}

To understand how NSC properties relate to their host galaxies, we require estimates of the stellar and total host galaxy masses.
We estimate the stellar mass ($M_{*,gal}$) using the stellar mass-to-light ratio calculated with the color-dependent transformations of \cite{roediger} and the measured luminosity and color of each component. We sum up the contributions from the S\'ersic components to estimate the stellar mass of the galaxy. Consequently, our values of $M_{*,gal}$ exclude the stars within the NSC. We can estimate masses only for systems with a measured or estimated distance.

We measure the stellar mass of the NSCs ($M_{NSC}$) using the same method employed for measuring the host galaxy stellar mass. Uncertainties in both the NSC luminosity and color are generally higher than those of the underlying galaxies because the measurements are highly sensitive to uncertainties in the PSF. Despite implementing a strict color uncertainty threshold for NSC photometry (discussed in \S\ref{sec:classification}), the mean error in NSC stellar mass is likely to be $\sim$ 0.1 dex. This estimate does not include systematic uncertainties arising from the choice of the mass-to-light calibration.

We estimate the halo mass ($M_{h,gal}$) using the photometric halo mass estimator proposed by \cite{2023MNRAS.519..871Z}, which relies on the galaxy's color, size, and luminosity. Because that method is calibrated for single S\'ersic models, we fit a single S\'ersic model, irrespective of the classification of the target galaxy, masking all identified point sources, including NSCs.
We calculate $M_{*,gal}$ and $M_{h,gal}$ only for those galaxies not in the `failed fitting' category. The halo mass estimator is only applicable to galaxies for which a solution to the scaling relation exists to provide an estimate of the velocity dispersion at $r_e$ and those are dark matter dominated within $r_e$. These requirement leads us to exclude 125 SMUDGes and 29 SDSS galaxies from further discussion when considering $M_{h,gal}$.

\subsection{Environment}
\label{sec:environment}

We adopt the distance to the 10th nearest galaxy in projection, referred to as D$_{10}$, as a measure of the galactic environment. To calculate D$_{10}$, we extract a magnitude-limited sample ($-14 \le $ M$_r/$mag $ \le -17$) of galaxies that have a recessional velocity within $\pm$ 1500 km s$^{-1}$ of our target \citep[roughly 2 times the velocity dispersion of the main component of the Virgo cluster;][]{binggeli} from SDSS,
sort those by projected separation from our galaxy, and evaluate the physical distance in projection to the 10th nearest neighbor. 
However, for some galaxies located at the boundary of the SDSS region or beyond, we are unable to measure the D$_{10}$ parameter.
To ensure that our search for nearby neighbors lies entirely within the SDSS footprint \citep{2020ApJS..249....3A}, we constrain our measurement of D$_{10}$ to targets
located an RA range of 120\textdegree $\leq \alpha \leq$ 250\textdegree\ and a Dec range of 0\textdegree $\leq \delta \leq$ 60\textdegree. Low values of D$_{10}$ correspond to high-density environments.

\section{Results}
\label{sec:results}

We explore two fundamental questions: 1) are the host galaxies of NSCs different from the parent sample of galaxies, and 2) are the properties of NSCs dependent on the host galaxy properties.  
We first examine how the morphology, structure, and environment of a host galaxy affect the likelihood of hosting an NSC. We then examine how the mass of the NSCs depends on host properties.

\subsection{Host Galaxy Properties}
\label{sec:results_gal}

Previous studies have already shown that NSC-bearing galaxies are distinct in stellar mass and environment. We now explore these and other possible relationships.

\begin{figure}
	\includegraphics[width=\columnwidth]{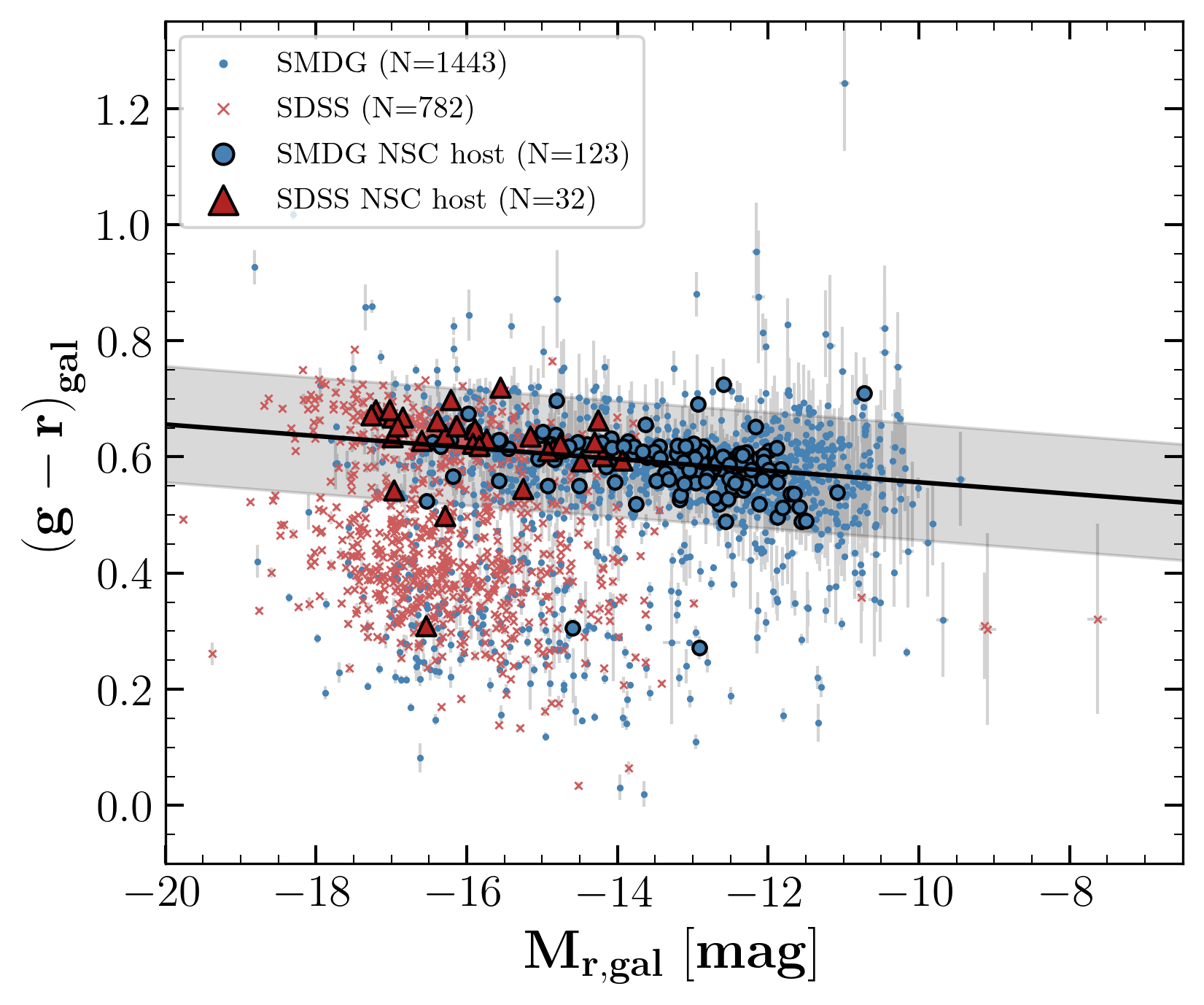}
    \caption{Color-magnitude relation for SMUDGes (blue) and SDSS (red) galaxies. Galaxies with NSCs are highlighted using circles (SMUDGes) and triangles (SDSS) and are almost exclusively found along the red sequence. The black line represents the fitted red sequence. Galaxies whose color difference from the red sequence is less than 0.1 mag are considered to be red sequence galaxies (shaded region). All data points have error bars representing the uncertainty measured from GALFIT.}
    \label{fig:cmd}
\end{figure}

\begin{figure}
	\includegraphics[width=\columnwidth]{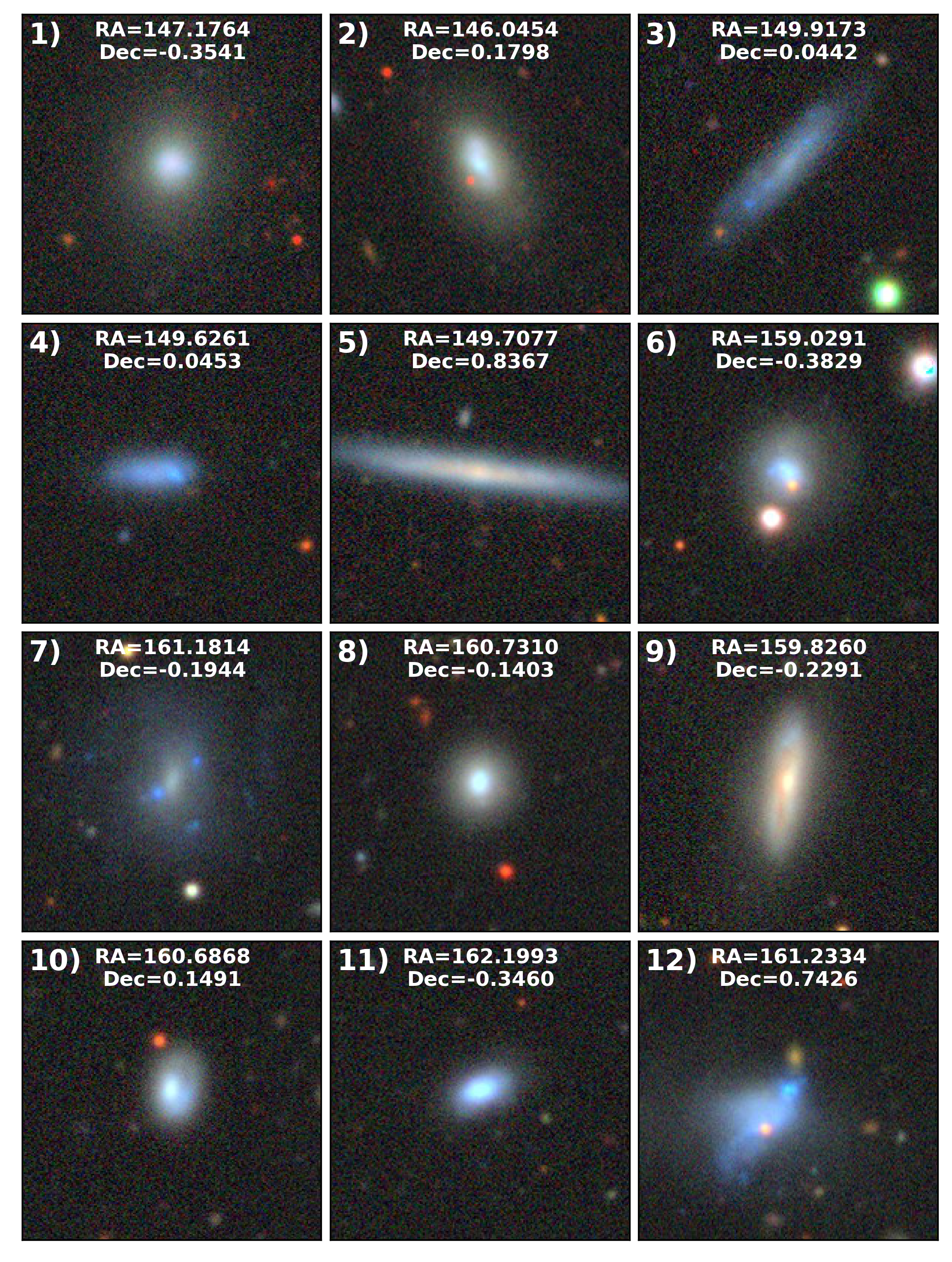}
    \caption{Cutout images of galaxies exemplifying cases where no satisfactory model was identified (referred to as the ``failed fitting" category). We display the first 12 such galaxies as sorted by SDSS `SpecObjID'. There is a range of morphology, although blue and highly inclined systems are overrepresented in the category.
    The field of view of each image is 52.4 arcsec. These images are drawn from the Legacy Surveys online viewer (\url{ https://www.legacysurvey.org/viewer}).
    }
    \label{fig:failed}
\end{figure}

\subsubsection{Galaxy Type}
\label{sec:morp}

Galaxy color is a broad measure of stellar populations and a proxy for morphology, particularly at this galaxy mass range where the S\'ersic index is almost exclusively $\sim 1$. In Figure \ref{fig:cmd} we compare the colors of galaxies with and without NSCs (excluding ``failed fitting" galaxies).
The NSC-bearing galaxies predominantly align along the galactic red sequence ($g-r \sim$ 0.6), while the hosts without an NSC exhibit a broad distribution of colors, characteristic of the general population.
We fit the red sequence (the black solid line in Figure \ref{fig:cmd}) using sigma clipping with a 2-sigma cut and 5 iterations. Subsequently, we define galaxies as red sequence galaxies if they lie within 0.1 mag in color from the fitted red sequence line (the shaded region). The majority of SMUDGes galaxies fall within the red sequence (890 out of 1151), with a higher fraction of NSC-bearing SMUDGes falling within the red sequence (116 out of 121).
There are only three truly blue ($g-r \le 0.4$) NSC hosts in our sample. 

We draw several conclusions from the Figure. 
The narrowness and tilt of the red sequence attest to our ability to measure galaxy colors even for our faintest galaxies. Furthermore, we find a qualitative difference in the color distribution of galaxies in the SDSS and SMUDGes samples. The fainter SMUDGes sample consists predominantly of red galaxies, as discussed by \cite{smudges5}, so at faint magnitudes the lack of NSCs in blue hosts could simply reflect the lack of blue faint galaxies, but not so for the SDSS sample. Finally, as already stated, our sample of NSC-bearing galaxies lies almost exclusively along the red sequence.

We do not infer from the last of these findings that only red galaxies host NSCs. A majority ($\sim 81$\%) of SDSS targets are classified as ``failed fitting", in contrast to the SMUDGes sample where $<6$\% are. Because we make no designation in ``failed" galaxies regarding the presence of an unresolved source, we may be missing many NSC-bearing galaxies in the SDSS sample. 
We attribute the difference in classification between the SDSS and SMUDGes samples to the inclusion in the SDSS sample of galaxies with highly non-uniform structures (e.g., spiral arms, star-forming regions) and galaxies with higher inclination. 
The ``failed fitting" galaxies are biased blue ($\langle g-r \rangle = 0.39$) and many are also highly inclined or asymmetric (Figure \ref{fig:failed}).
Even though GALFIT results mostly converge for these more complex galaxies, we have rejected these models because they are statistically unacceptable on goodness-of-fit grounds.

We conclude that our following results pertain only to red or quiescent galaxies and to NSCs in such galaxies, and that we may be undercounting NSCs in galaxies that are not on the red sequence. Thus, we are not in a position to address questions of NSC properties across host galaxy types as measured by color.

\begin{figure}
	\includegraphics[width=\columnwidth]{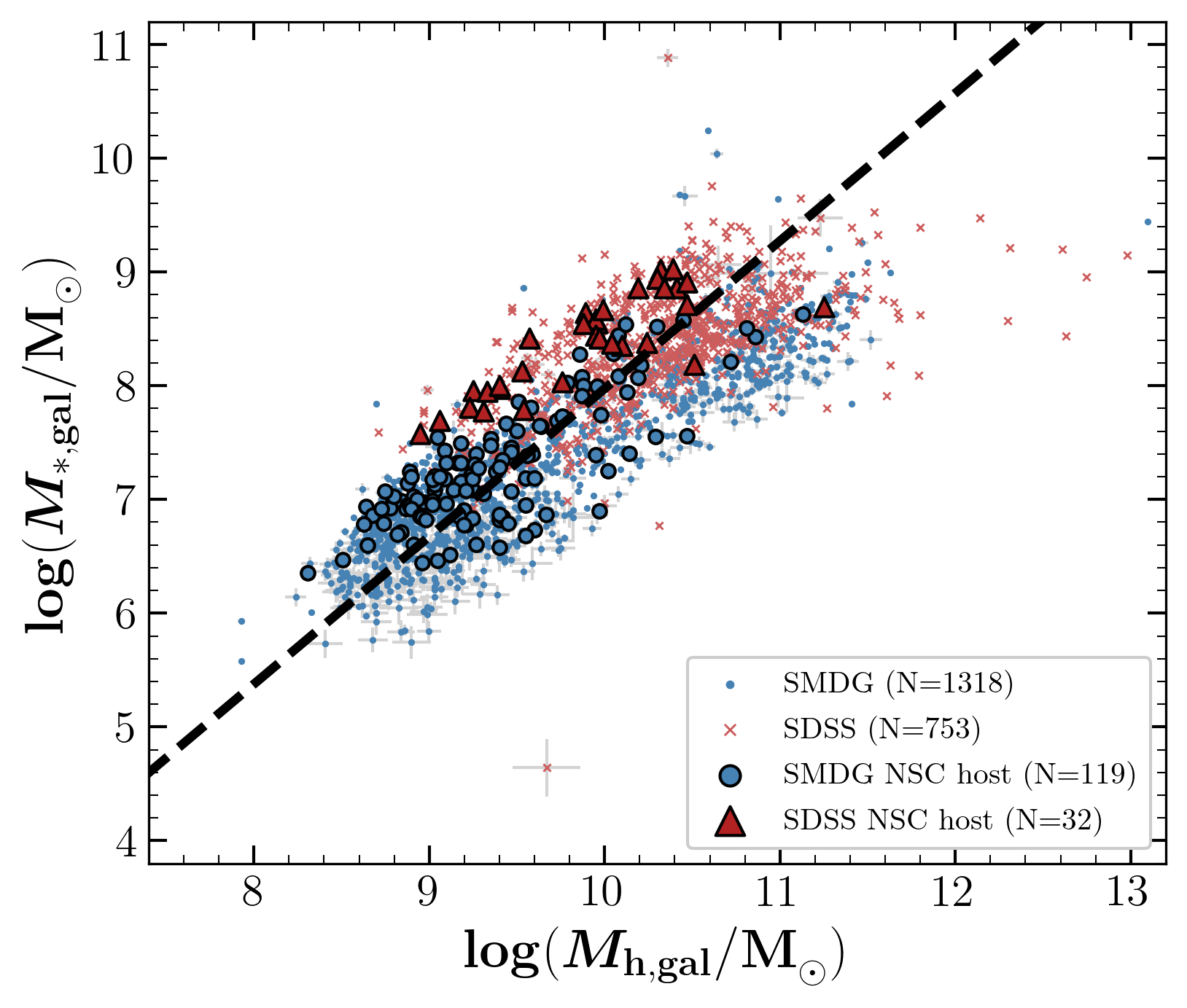}
    \caption{Stellar and estimated halo mass of SMUDGes (blue) and SDSS (red) galaxies. Galaxies with NSCs are highlighted using circles (SMUDGes) and triangles (SDSS) and are, in general, biased upward from the parent galaxy population. 
    }
    \label{fig:mass_hmass}
\end{figure}

\subsubsection{Galaxy Structure}

We present the stellar and halo masses of our galaxies in Figure \ref{fig:mass_hmass}.
The galaxies densely sample stellar masses from $\sim$ $10^{6.5}$ \Msol\ to $10^9$ \Msol. The median SMUDGes galaxy stellar mass  ($10^{7.1}$ \Msol) is almost a factor of ten lower than that of an SDSS galaxy ($10^{8.0}$ \Msol), although the samples overlap significantly. 
The galaxies densely sample estimated halo masses from $\sim 10^{8.5}$ \Msol to $10^{11.5}$ \Msol, again with significant overlap. 

A principal difference between the two samples, which is not evident in the Figure, is in the central surface brightness ($\mu_{0,g}$).
The $g$-band central surface brightnesses of SMUDGes galaxies are, by selection, fainter than 24 mag arcsec$^{-2}$ (See \S \ref{sec:data}). In contrast, there is no specific central surface brightness cut for SDSS galaxies, although the requirement for a high S/N spectroscopic measurement is likely to lead to a bias toward much lower (brighter) central surface brightness. The median surface brightness for SDSS galaxies in our sample is 21.2 mag arcsec$^{-2}$, whereas that of SMUDGes is 24.76 mag arcsec$^{-2}$. Extending the central surface brightness range for our study was a principal motivation for adding the SDSS sample to the SMUDGes sample.

We find a clear offset between the NSC-bearing galaxies and the full galaxy population in the $M_{*,gal}$-$M_{h,gal}$ space. The NSC-bearing galaxies can be described as either having a larger stellar mass for a given halo mass, or a lower halo mass for a given stellar mass. Quantitatively, the NSC-bearing galaxies have $\langle \log_{10}$($M_{*,gal}$/$M_{h,gal})\rangle$ values of $-$2.08 $\pm$ 0.03 and $-$1.41 $\pm$ 0.05 (SMUDGes and SDSS, respectively), which are significantly larger than the corresponding values for the non-bearing galaxies, $-$2.29 $\pm$ 0.01 and $-$1.99 $\pm$ 0.02.
We will revisit this difference when discussing occupation fractions in \S\ref{sec:occupation_fractions}.

\begin{figure*}
	\includegraphics[width=\textwidth]{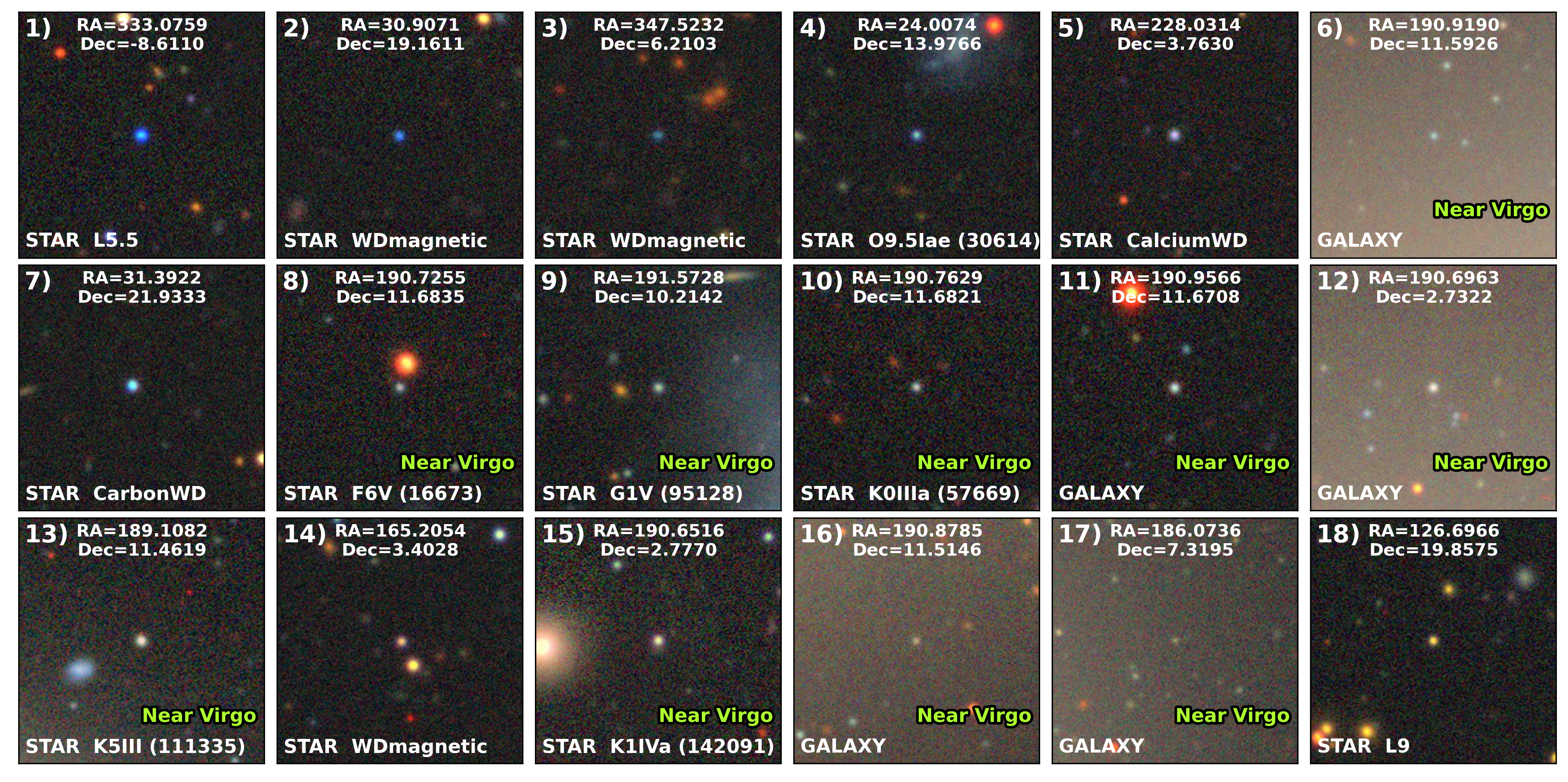}
    \caption{Cutout images of 18 bare PSF (PSF only) targets in our SDSS sample sorted by increasing $g-r$. The field of view of each image is 52.4 arcsec. The SDSS spectroscopic classification is provided at the bottom of each panel. We also indicate if the target is located within 10 degrees of the Virgo cluster center. These images are drawn from the Legacy Surveys online viewer (\url{ https://www.legacysurvey.org/viewer}).}
    \label{fig:bare_psf}
\end{figure*}

\subsubsection{Absent Hosts?}
\label{sec:bare_psf}

Our modeling identifies 18 SDSS sources that have a sufficiently large recessional velocity (cz $>$ 1000 km s$^{-1}$) to place them well beyond the anticipated range of Galactic stars but which are nevertheless unresolved and apparently not associated with any underlying diffuse stellar component. We present these in Figure \ref{fig:bare_psf}, ordered in increasing $g-r$. The absolute $r$-band magnitudes fall within a narrow range of $-11.5$ to $-9.1$ mag, centered within the absolute magnitude range of our hosted NSCs, and their color ranges from $-$0.5 to 1.5.

These ``Bare PSFs'' can be divided into several groups.
Among these 18 objects, 10 are located within the Virgo cluster region (i.e., within 10 degrees from the Virgo cluster center). Six of these (objects 6, 9, 12, 13, 16, and 17 in Figure \ref{fig:bare_psf}) are projected on the stellar halo of a nearby giant galaxy and also have a redshift that is similar to that of the nearby giant.
The remaining four targets near Virgo do not have an obvious association with a Virgo galaxy. Objects 1 through 5 are extremely blue $(g-r <0.4$), and the SDSS spectroscopic pipeline classified 3 of 5 as white dwarf stars. Despite their large recessional velocities, these are likely galactic stars. \cite{2023OJAp....6E..28E} present white dwarfs with radial velocities $>$ 1000 km s$^{-1}$ that are thought to be runaways from Type Ia supernovae. Objects 7 ($g-r =0.60$) and 14 ($g-r =0.75$), which are relatively redder, are also spectrally classified as white dwarfs and may belong to this class of runaways as well. In the end, there is only one object (Object 18) that is isolated (not in the Virgo region) and not spectroscopically classified as a white dwarf. Isolated NSCs are a rare phenomenon, if they exist at all, outside of dense environments. 

\begin{figure}
	\includegraphics[width=\columnwidth]{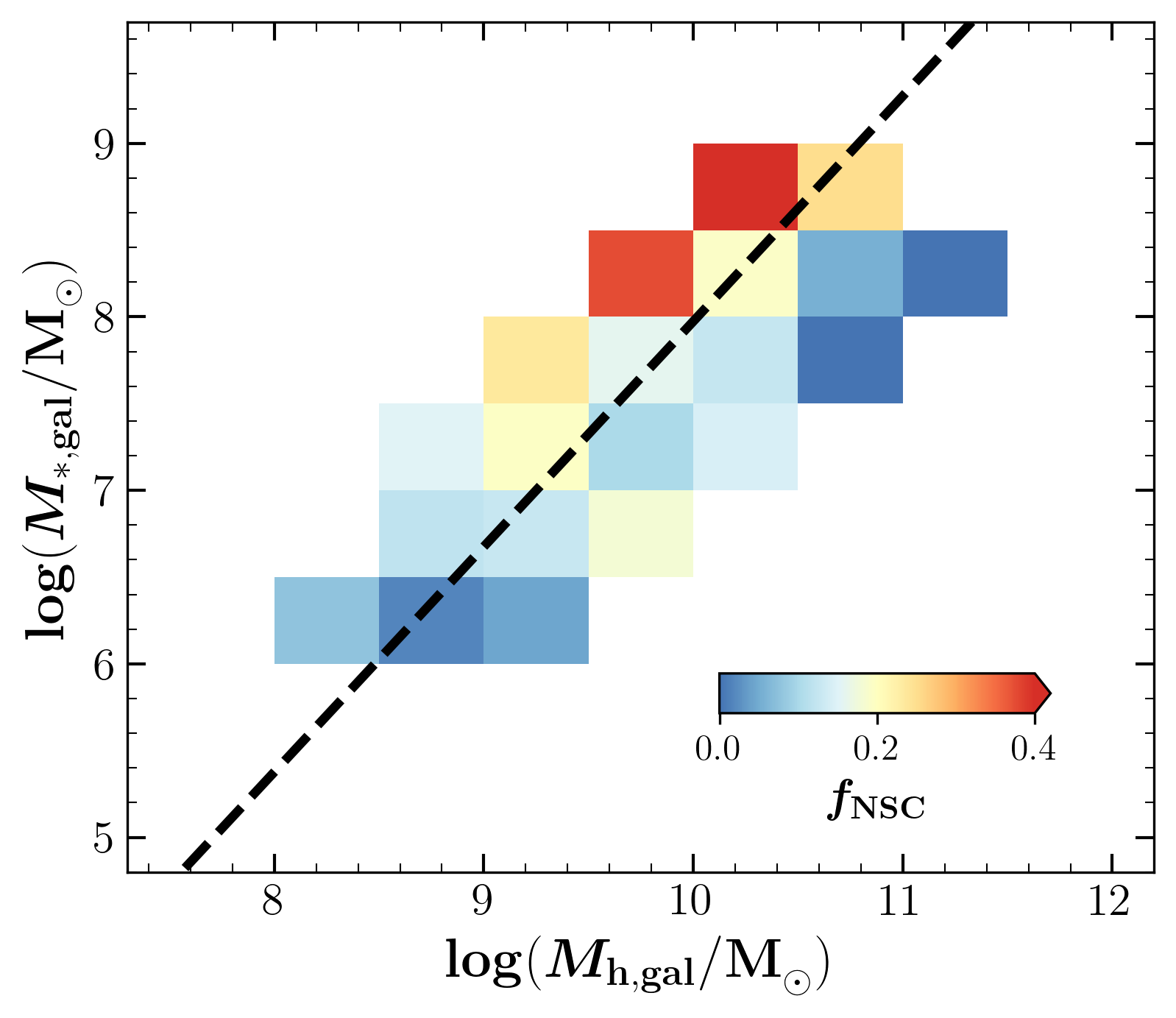}
    \caption{
    The dependence of NSC occupation fraction, $f_{NSC}$, on stellar and halo mass for the red sequence SMUDGes sample.  
    The dashed black line shows the mean $M_{*,gal}-M_{h,gal}$ relation from \cite{2023MNRAS.519..871Z} for reference. Each pixel with a plotted occupation fraction must contain at least 4 galaxies. Galaxies above the mean stellar mass-halo mass relation have higher $f_{NSC}$ than those below the line.
    }
    \label{fig:smdgdist_nsc2dfrac_MhM}
\end{figure}

\begin{figure}
	\includegraphics[width=\columnwidth]{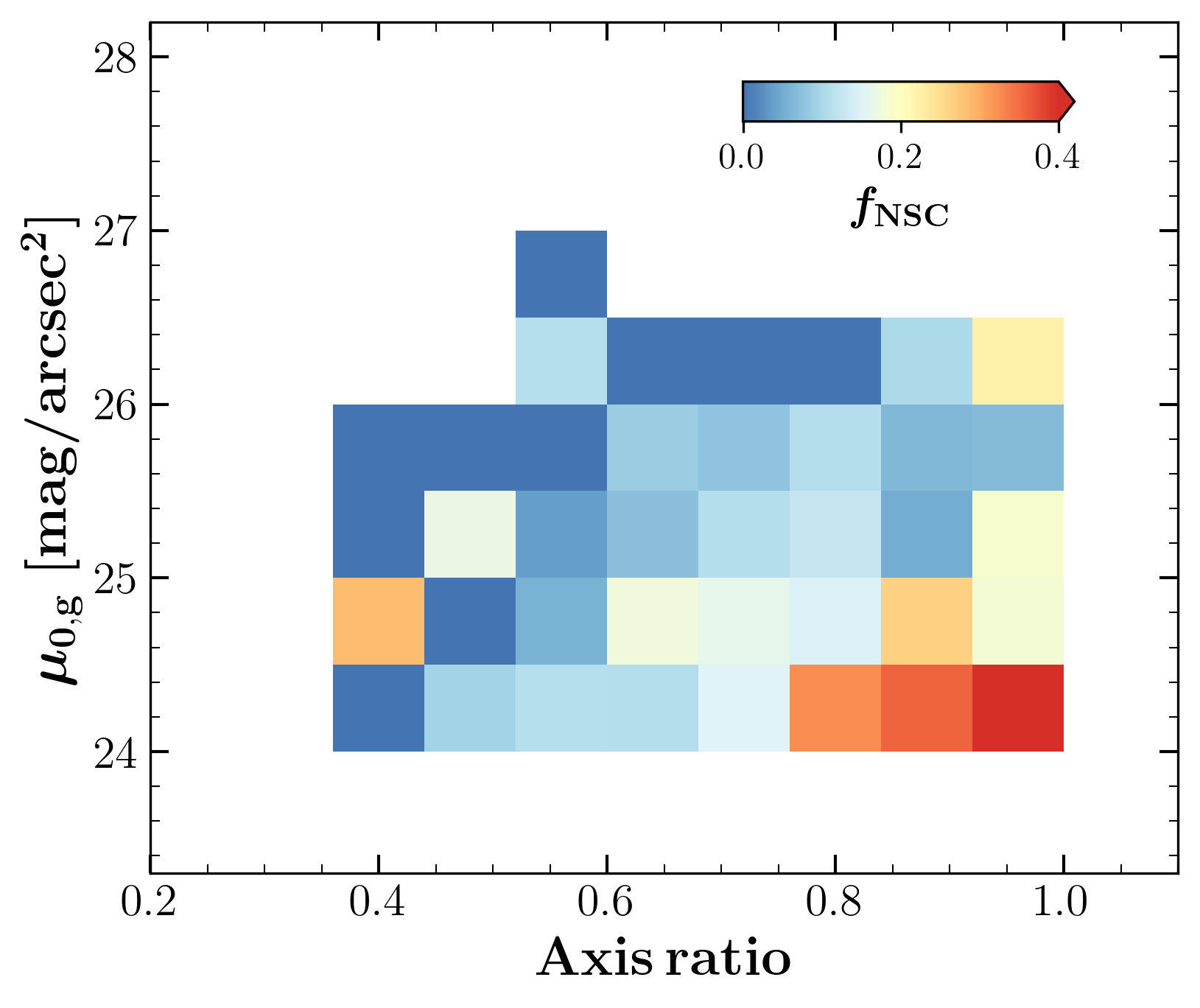}
    \caption{The dependence of the NSC occupation fraction, $f_{NSC}$, on central surface brightness and axis ratio for the red sequence SMUDGes sample.
    There is a decline in occupation fraction with fainter central surface brightness and with a smaller axis ratio.
    }
    \label{fig:smdgdist_nsc2dfrac_muAR}
\end{figure}

\subsubsection{NSC Occupation Fractions}
\label{sec:occupation_fractions}

In this subsection, we measure the fraction of our galaxies that host an NSC, the occupation fraction or $f_{NSC}$, and how $f_{NSC}$ varies across galaxies with different characteristics.  Because of the differences in our ability to identify NSCs in the SMUDGes and SDSS samples (see \S\ref{sec:morp}), we present results only for the larger SMUDGes sample in the Figures in this section unless otherwise noted, but also check to see whether the SDSS sample is consistent with any results drawn. That comparison is presented at the end of this section. 

A second precaution involves the color of the host galaxy. Because we find that nearly all NSC-bearing SMUDGes galaxies are on the red sequence, we focus our investigation of the dependence of $f_{NSC}$ on galaxy properties among red sequence galaxies. 
We do this to avoid blending the real or apparent $f_{NSC}$ color dependence (\S\ref{sec:morp}) into other relations. For example, one can imagine that a relation between $f_{NSC}$ and environment could arise simply from the combination of the relations between $f_{NSC}$ and galaxy color, and galaxy color and environment. Therefore, the following analyses use only the red sequence SMUDGes galaxies, those within the shaded region in Figure \ref{fig:cmd}. However, we find that the overall trends are the same when we instead use the entire SMUDGes sample.

We begin by examining how $f_{NSC}$ varies with $M_{*,gal}$ and $M_{h,gal}$ in Figure \ref{fig:smdgdist_nsc2dfrac_MhM}.
We find a higher occupation fraction in galaxies with higher stellar mass, as seen previously \citep{denBrok2014, sanchez2019, neumayer}, but the behavior of $f_{NSC}$ is clearly more complicated than a simple dependence on $M_{*,gal}$. More accurately, it appears that $f_{NSC}$ increases with increasing stellar mass fraction ($M_{*,gal}/M_{h,gal}$). Galaxies above the mean stellar mass-halo mass relation for galaxies of this mass (the black line, adopted from  \citep{2023MNRAS.519..871Z}) have on average a larger $f_{NSC}$ than those below
(as already suggested in Figure \ref{fig:mass_hmass}). %This finding is even clearer in the 1-D histogram presented in Figure \ref{fig:nsc1dfrac}. 

In Figure \ref{fig:smdgdist_nsc2dfrac_muAR} we show $f_{NSC}$ as a function of the central surface brightness and axis ratio of the host galaxy. Here we confirm that galaxies with a brighter central surface brightness, at least for the range probed by SMUDGes, tend to have a higher occupation fraction \citep{Lim2018}. %This population corresponds to the `high-stellar-fraction' galaxies we discussed above. 
Also from this Figure, we infer that rounder galaxies (higher axis ratio) tend to have a higher occupation fraction \citep{Lisker07, sanchez2019b}. We determine that the latter finding is not the result of varying incompleteness as a function of galaxy inclination or morphology using the completeness simulations described previously 
(\S\ref{sec:completeness}) by verifying that our estimate of the completeness does not depend on the axis ratio. Note that systems with smaller axis ratios may have intrinsically lower central surface brightness than measured if they are oblate systems viewed edge-on. As such, these two parameters may be related, depending on the intrinsic shape of these galaxies.

\begin{figure}
	\includegraphics[width=\columnwidth]{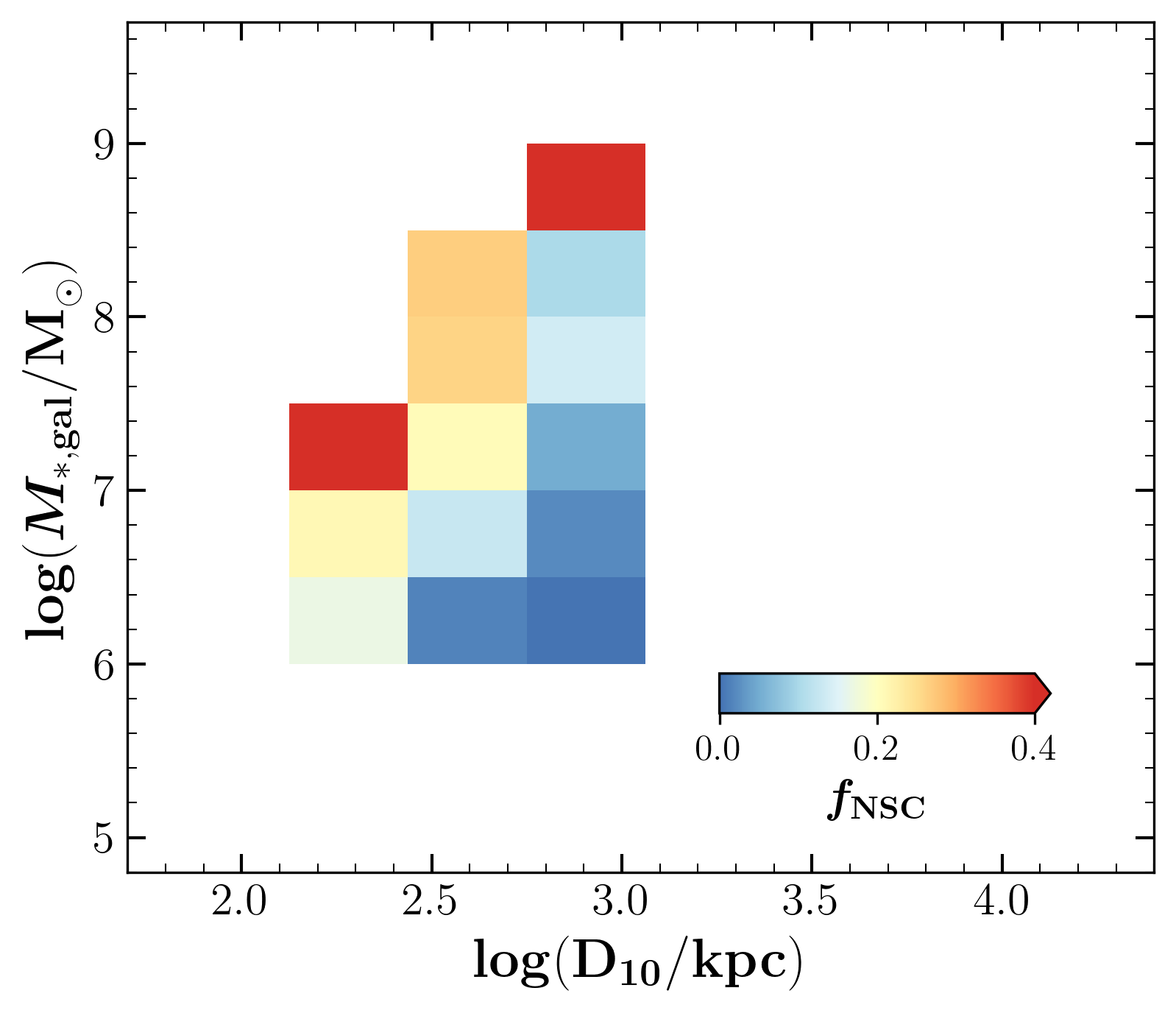}
    \caption{The dependence of NSC occupation fraction, $f_{NSC}$, on host stellar mass and the environmental measure D$_{10}$ for the red sequence SMUDGes sample. A dependence on both variables is evident.}
    \label{fig:paper_smdgdist_nsc2dfrac_MgEnv}
\end{figure}

Finally, in Figure \ref{fig:paper_smdgdist_nsc2dfrac_MgEnv} we show $f_{NSC}$ as a function of D$_{10}$, our environmental tracer (see  \S\ref{sec:environment}). Smaller values of D$_{10}$ correspond to higher-density environments. We find that $f_{NSC}$ varies both with $M_{*,gal}$, as seen before in Figure \ref{fig:smdgdist_nsc2dfrac_MhM}, and D$_{10}$. Both of these dependencies had been identified previously \citep[e.g.,][]{denBrok2014, sanchez2019, Lim2018, Carlsten_2022}.

\begin{figure}
	\includegraphics[width=\columnwidth]{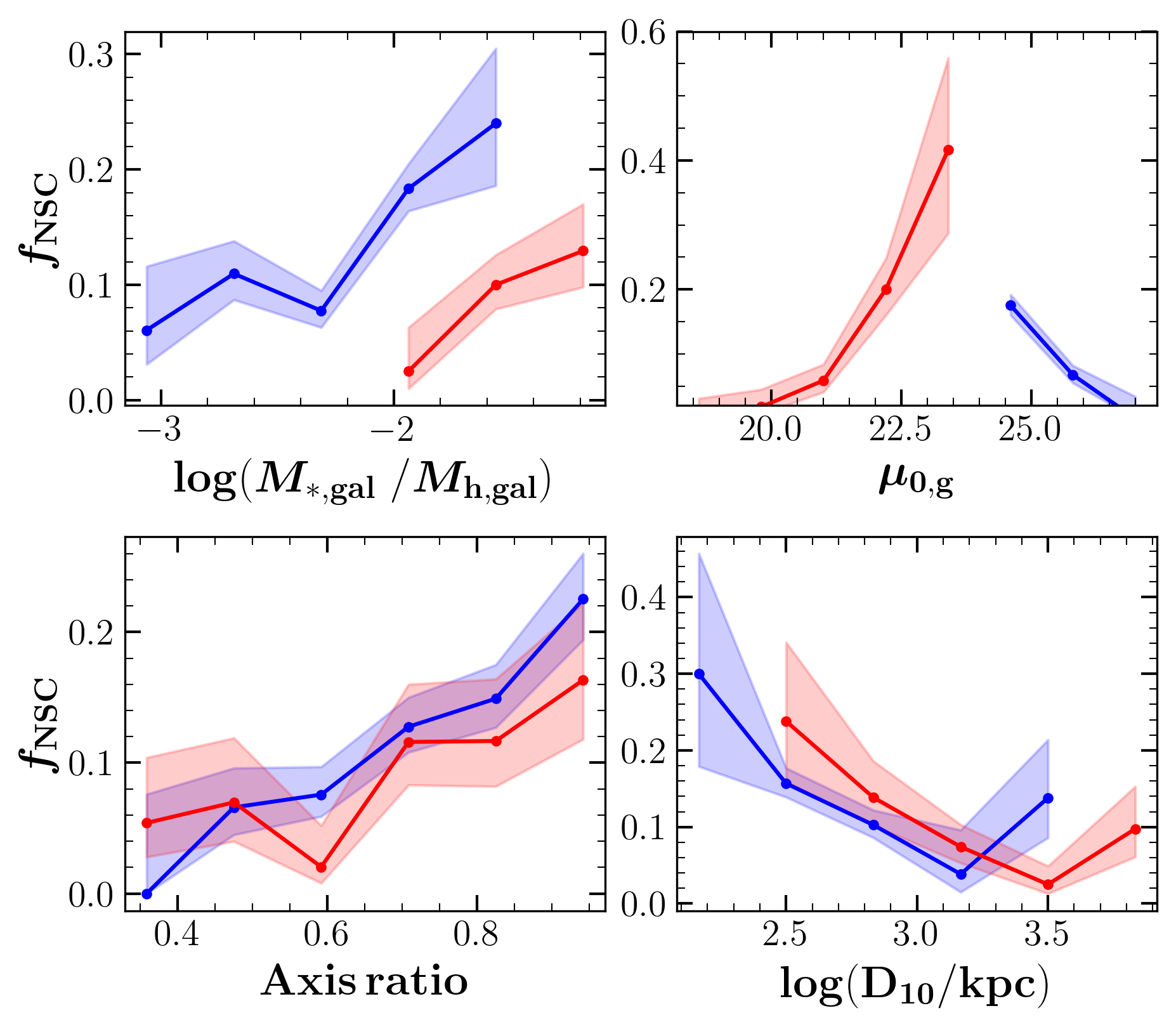}
    \caption{The dependence of NSC occupation fraction, $f_{NSC}$, on stellar mass fraction, surface brightness, axis ratio, and D$_{10}$ for the red sequence SMUDGes (blue) and SDSS (red) samples. 
    The error bars are calculated using the  Binomial distribution for the appropriate number of objects in each bin and represent the 1$\sigma$ uncertainties. The two samples show the same trends in $f_{NSC}$ where they overlap, although the normalizations may differ for reasons described in the text.
    }
    \label{fig:nsc1dfrac}
\end{figure}

As we described previously,
we utilized only the SMUDGes sample in the 2D histograms presented in this Section due to differences in NSC identification rates between the SMUDGes and SDSS samples. However, in Figure \ref{fig:nsc1dfrac} we now demonstrate that the trends in $f_{NSC}$ found in the red sequence SMUDGes sample match those from the red sequence SDSS sample when the parameter ranges overlap. A different normalization in $f_{NSC}$ between the samples is possible given our completeness differences.  The one reversal in behavior is in central surface brightness ($\mu_{0}$), although because the ranges do not overlap the results are not in direct conflict. We have somewhat less confidence in the SDSS result because the NSC incompleteness increases with increasing central surface brightnesses (see \S\ref{sec:completeness}), which is in the same sense as the measured trend for that sample. Overall, however, we find convincingly that $f_{NSC}$ correlates with a wide variety of host galaxy properties ($M_{*,gal}$,$\mu_0$, stellar mass fraction, axis ratio, and environment). We will return to the complication that many of these properties correlate with each other. 

\begin{figure}
	\includegraphics[width=\columnwidth]{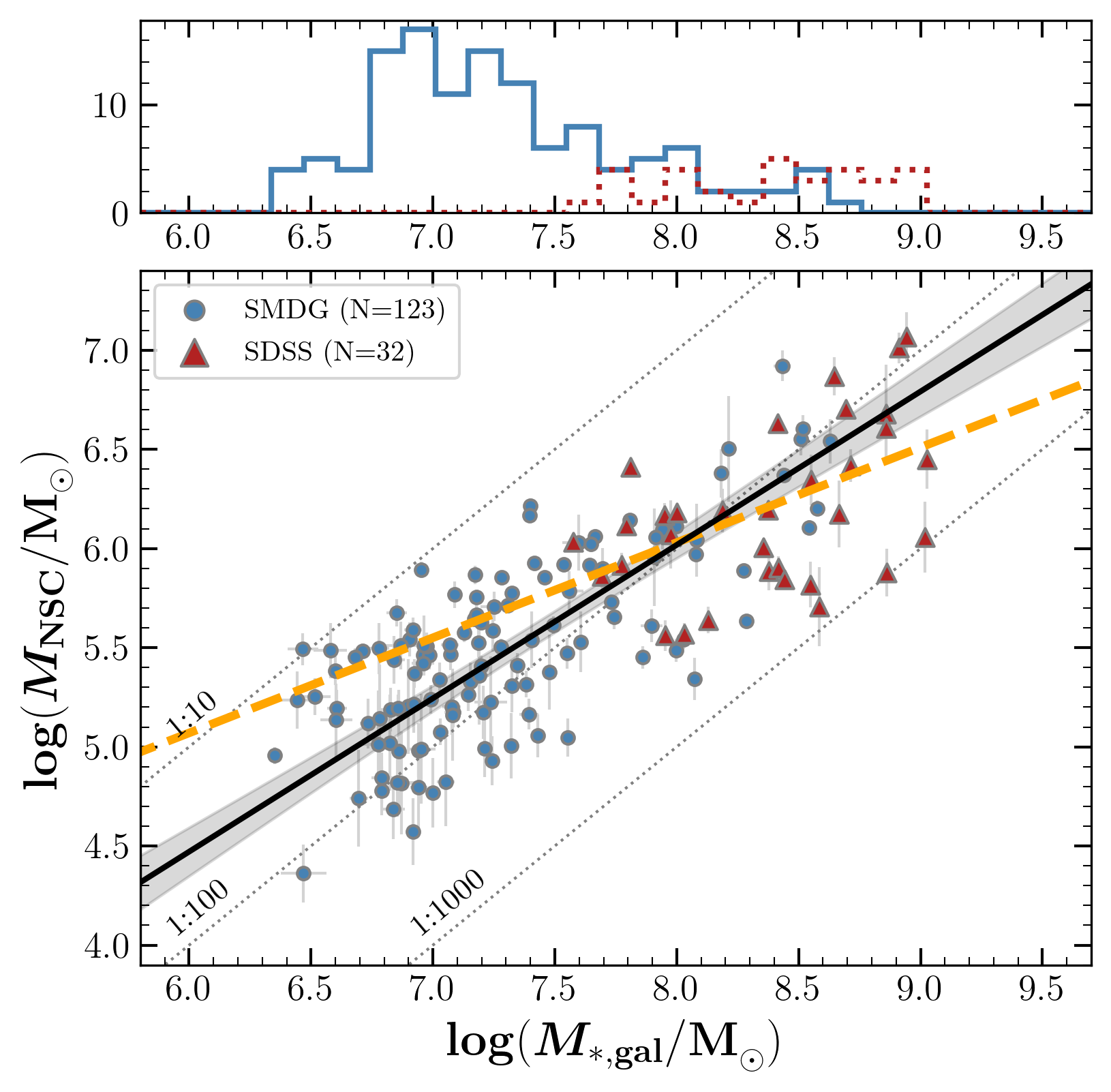}
    \caption{NSC stellar mass vs. host galaxy stellar mass for SMUDGes (circles) and SDSS galaxies (triangles). 
    The stellar mass distribution of the SMUDGes (solid line) and SDSS (dotted line) galaxies are in the upper panel.
    In the lower panel, the solid line and shaded region represent the best-fit line and its 1$\sigma$ uncertainty. For comparison, we also present the relationship from \cite{neumayer} (dashed line) and lines of proportionality for three different fractional levels (dotted lines). We find $M_{NSC}/$\Msol$ = 10^{6.02\pm0.03}(M_{*,gal}/10^{8} $\Msol$)^{0.77\pm0.04}$ with 0.27 dex scatter.}
    \label{fig:both_mass_mass}
\end{figure}

\subsection{NSC Properties}
\label{sec:results_nsc}

We have established that the likelihood of a galaxy hosting an NSC is a function of various properties of the host galaxy. Now we address whether the masses and colors of NSCs depend on host galaxy properties.

\subsubsection{The $M_{NSC}$ -- $M_{*,gal}$ Scaling Relation}

We present the relationship between the stellar mass of the NSC ($M_{NSC}$) and host galaxy ($M_{*,gal}$) in Figure \ref{fig:both_mass_mass}. Our two samples appear to follow the same relationship, with the SDSS sample providing better sampling at higher stellar masses and SMUDGes at lower masses. 
A scaling of $M_{NSC}$ with $M_{*,gal}$ is clear, but determining the relationship has subtleties that are not always appreciated. 
The gray dotted lines in the Figure display proportionality for three different fractional levels at 1 dex intervals. The fraction of stellar mass that is in the NSCs ranges from 0.1 to 0.001. It is striking that there are systems for which nearly 10\% of the stellar mass is in the central cluster.

Fitting linear relationships to data with significant scatter, whether that scatter is real or observational, is a delicate matter \citep{isobe}. While there is no definitively superior approach, \cite{isobe} conclude that the most robust approach is that referred to as the bisector ordinary least squares method, which is different than the ordinary least squares method that is generally used. 

Our fit using the bisector ordinary least square method results in the relationship $M_{NSC}/$\Msol$ = 10^{6.02\pm0.03}(M_{*,gal}/10^{8} $\Msol$)^{0.77\pm0.04}$ with a scatter of 0.27 dex.
If we instead use the traditional ordinary least square method (i.e., OLS(Y|X)), the result is $M_{NSC}/$\Msol$ = 10^{5.95\pm0.04}(M_{*,gal}/10^{8} $\Msol$)^{0.64\pm0.04}$.
The derived slopes differ significantly even though they are measured from the same data. This dependence on fitting methodology highlights one difficulty in comparing results among studies.

\begin{figure}
	\includegraphics[width=\columnwidth]{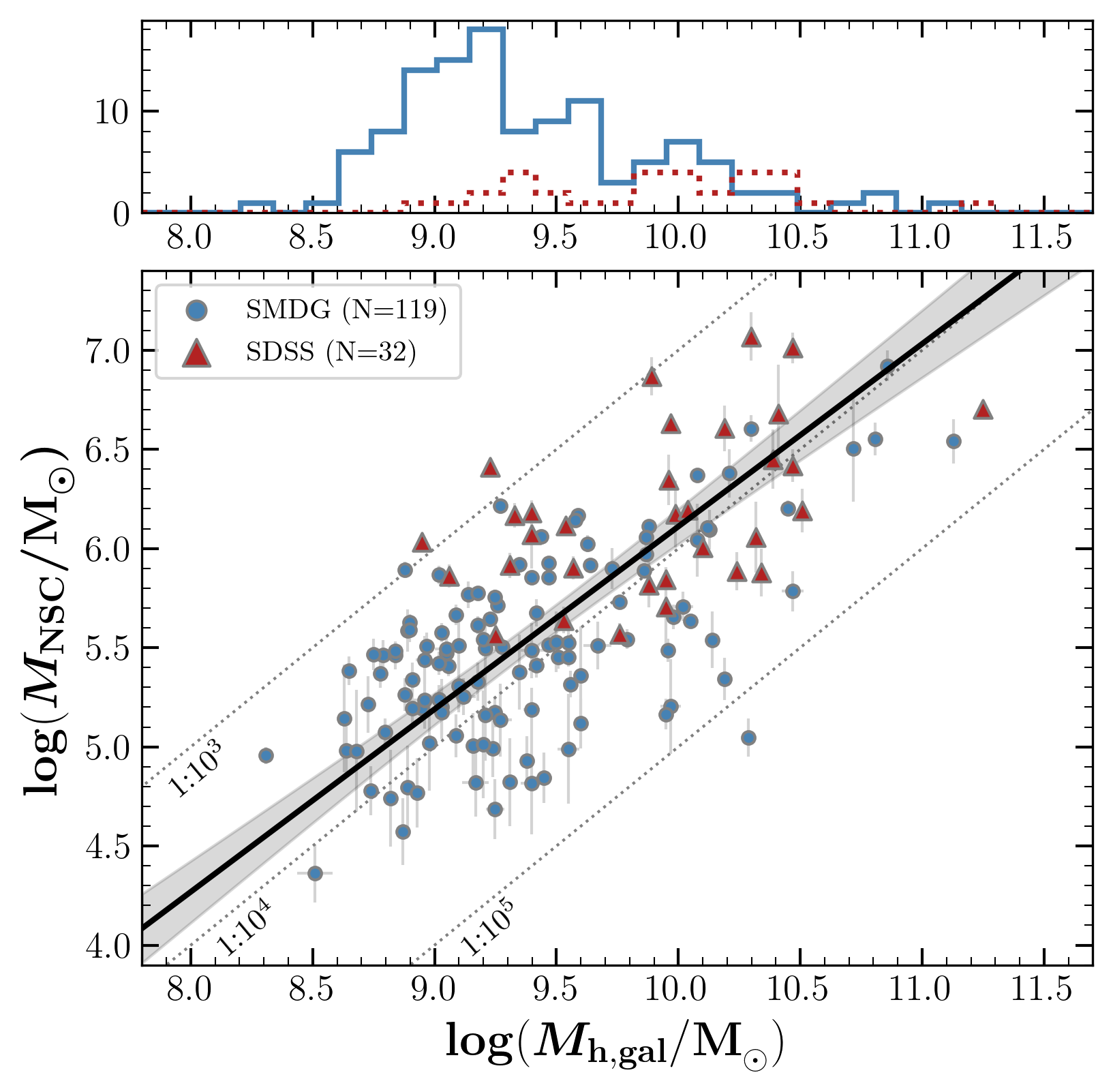}
    \caption{NSC stellar mass vs. host galaxy halo mass in the same format as in Figure \ref{fig:both_mass_mass}. 
    We find $M_{NSC}/$\Msol$ = 10^{6.11\pm0.05}(M_{h,gal}/10^{10} $\Msol$)^{0.92\pm0.05}$ with 0.32 dex scatter.}
    \label{fig:both_mass_hmass}
\end{figure}

As an example of such a comparison, we consider results from \cite{neumayer}. Because our galaxies are predominantly red (Figure \ref{fig:cmd}) and the stellar mass range is comparable to the early-type galaxies in \cite{neumayer}, we compare our results to their early-type galaxy relation (orange dashed line in Figure \ref{fig:both_mass_mass}). 
Their measured slope of $0.48\pm0.04$ is more than 5.5 sigma discrepant with our preferred fit, but only 2.7 sigma discrepant if we fit using the ordinary least squares method. Upon further examination, we find that they also implement a lower limit on $M_{NSC}$. When we apply a similar mass-cut (rejecting log$(M_{NSC}/ $\Msol$) < 5$) and fit using the ordinary least squares method that they used, we obtain a slope of $0.63\pm0.04$, which is only slightly less discrepant. We conclude that the fitting method plays a significant role in comparing across samples, but that even using a consistent approach we have a mild discrepancy ($< 3\sigma$) with the result presented by \cite{neumayer}
(and the references therein from which they collected data). 
This discrepancy may reflect procedural differences, such as how the stellar masses are calculated, or actual physical differences, such as a slightly different $M_{NSC}$-$M_*$ relation for low surface brightness galaxies, which are more highly represented here. Indeed, a recent study found a much steeper slope of $0.82\pm0.08$ for field dwarfs \citep{hoyer23}, suggesting that a mixed environment sample such as ours would find an intermediate slope value.
Regardless, in the qualitative finding that the relationship between $M_{NSC}$ and $M_{*,gal}$ is sublinear there is full agreement.

\subsubsection{The $M_{NSC}$ -- $M_{h,gal}$ Scaling Relation}

We present the relationship between $M_{NSC}$ and $M_{h,gal}$ in Figure \ref{fig:both_mass_hmass}.
Again using the preferred bisector ordinary least squares method, we find $M_{NSC}/$\Msol$ = 10^{6.11\pm0.05}(M_{h,gal}/10^{10} $\Msol$)^{0.92\pm0.05}$ with a scatter of 0.32 dex.
The fitted slope suggests that $M_{NSC}$ is directly proportional to $M_{h,gal}$. 
The mean mass NSC fraction is $\sim$ $10^{-4}$, or alternatively stated 0.01\% of the mass of a galaxy resides in the NSC, to within a scatter of roughly a factor of 2, for galaxies that host an NSC. 

\begin{figure}
	\includegraphics[width=\columnwidth]{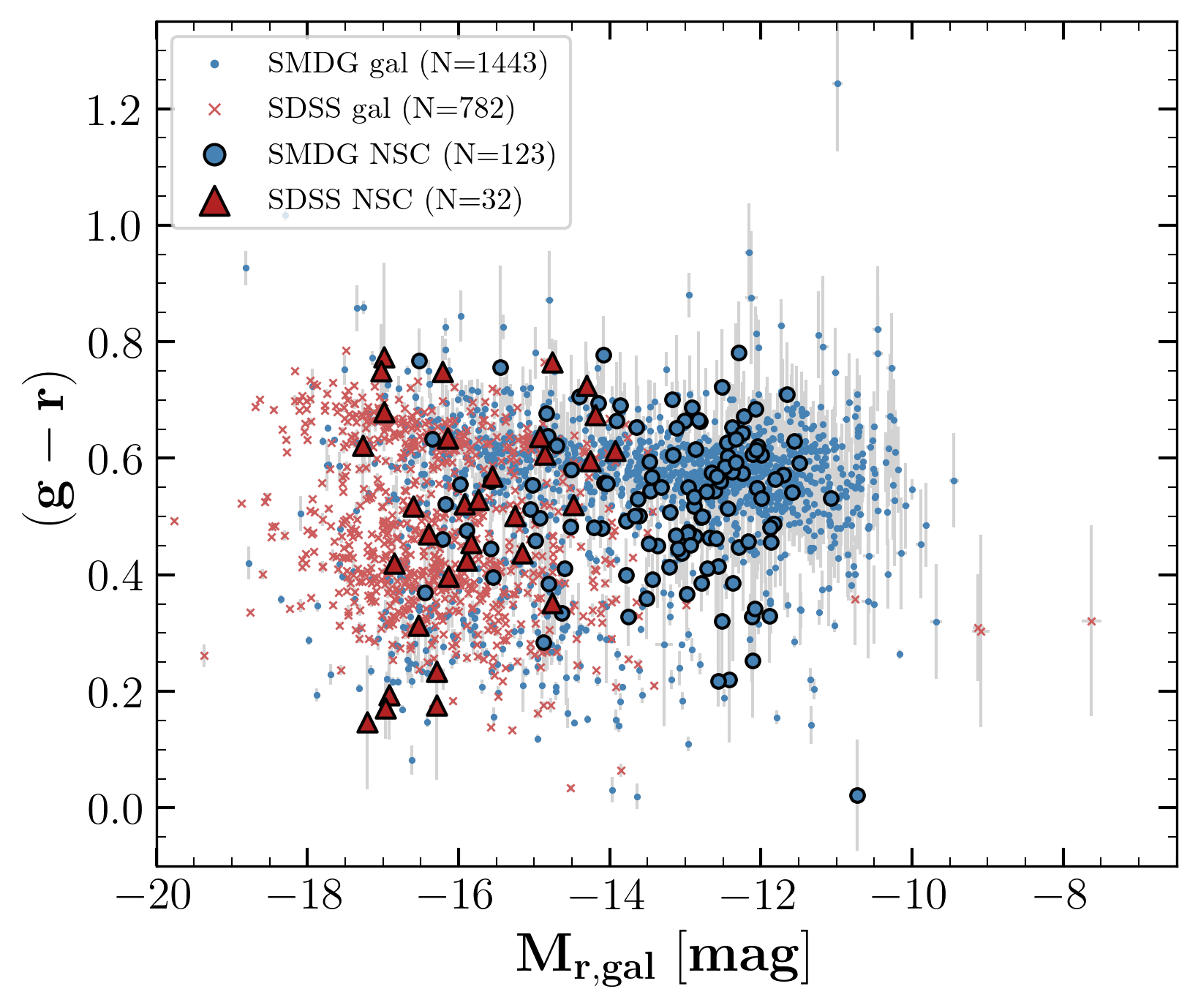}
    \caption{NSC Color vs. galaxy magnitude diagram in the same format as Figure \ref{fig:cmd}. Markers with a black edge show the NSC color as a function of the absolute magnitude of the host galaxy. For a comparison, we plot the ordinary color-magnitude diagram with small dots.}
    \label{fig:cmd_nsc}
\end{figure}

\subsubsection{NSC Color}
\label{sec:results_nsc_color}

We find that the color of the NSC shows a far broader distribution than that of the host galaxy (Figure \ref{fig:cmd_nsc}). However, we caution that the NSC colors are more difficult to measure. 
To test our NSC color measurements, we re-measure the colors using the 10th data release (DR10) of the Legacy Survey. For nearly a quarter of the NSC-bearing galaxies in the SMUDGes sample (27 out of 123), the measured NSC color changes by more than 0.2 mag. In contrast, the host galaxy color changes by this much for only one. %We attribute the differences in NSC colors to subtle changes in the model PSF between the bands.
We conclude that these images, and the associated PSFs, are insufficient to produce robust NSC color measurements. 

Even so, it is particularly interesting that there are some NSCs for which we measure a blue color. Given our skepticism of the measured colors, we checked GALEX photometry and 
the SDSS spectra of the galaxies hosting the blue NSCs for signs of recent or ongoing star formation, but do not find any. Despite these issues, we do find that the median NSC color is slightly, but significantly, bluer than that of their host ($g-r = 0.531\pm0.011$ vs. $0.599\pm0.005$, respectively). This result is consistent with the findings of \cite{sanchez2019, Carlsten_2022}.
Further study is required to confirm and explore the NSC colors, which when measured precisely will aid greatly in discriminating among formation models.

\subsection{Distance Uncertainties}

We conclude our presentation of the results by returning to a thorny issue.
While all of the SDSS galaxies and some SMUDGes have spectroscopically measured redshifts, we rely on the estimated distances for approximately 85\% of SMUDGes galaxies in our sample. An inaccurate distance estimate will significantly impact the calculated physical size, absolute magnitudes, stellar mass, and halo mass of the galaxy.
\cite{smudges5} found that the estimated redshifts were accurate about 70\% of the time by comparing estimated distances to spectroscopically measured ones. 

Nevertheless, we have reasons to be somewhat more confident of our estimated distances.
Presently, we have 21 additional galaxies with both spectroscopic measured redshift and estimated redshift, and only one shows a significant difference. 
Moreover, we expect that the uncertainty in the estimated distances will have a diminished impact on NSC scaling relations because the uncertainties are greatest in SMUDGes galaxies that are large and blue \citep[\S4.1 in][]{smudges5}, which are not the NSC-bearing galaxies in our sample. Finally, we evaluate the slopes in the scaling relationships using only host galaxies with spectroscopic redshifts. For the $M_{NSC}-M_{*,gal}$ relation we find an almost identify slope, $0.75 \pm 0.07$ vs. our original $0.77 \pm 0.04$. For the $M_{NSC}-M_{h,gal}$ relation we find a larger change relative to the full sample result, but the resulting slope is still within 2.3$\sigma$ of a direct proportionality, $0.82\pm0.08$.

We conclude that inaccurate distances, which will be in the minority, will contribute to the measured scatter in our relationships, but that the qualitative results are likely to remain unchanged. This claim is further supported by the consistency with the SDSS sample results, which are based entirely on spectroscopic distances. Moreover, because surface brightness, axis ratio, and the stellar mass fraction (distance affects stellar and halo masses similarly) are distance independent, 
the relationships between occupation fraction and these factors (e.g., Figure \ref{fig:smdgdist_nsc2dfrac_muAR}) are unaffected by distance errors. Even so, obtaining spectroscopic redshifts of galaxies in this sample with NSCs is a high priority, particularly for those that may be the most scientifically interesting such as those with the largest NSC mass fractions.

\section{Discussion}
\label{sec:discussion}

We now discuss some of our findings within the context of the two families of formation scenarios. 

\subsection{Are NSCs Predominantly in Early-Type Galaxies?}
\label{sec:discuss_morphology}

One potential difference between the two sets of NSC formation scenarios concerns their impact on morphology. The GC inspiral model would naturally generate NSCs across various morphological types, as the number of GCs depends on mass rather than morphology. On the other hand, the extreme in-situ star formation scenario might imply a significant dependence on morphology if the type of merger event proposed as the trigger for the star formation episode influences morphology.

\cite{2020MNRAS.491.1901H} used the MATLAS survey and suggested that early types have a higher occupation fraction. 
Our results appear to confirm this trend, 
but we discussed how our sample suffers from at least two biases against finding NSCs in blue galaxies. First, our fitting method rejects galaxies with complex structures that are not fitted by single or double S\'ersic components. Notably, most of the ``failed fitting" galaxies in our data are located in the blue cloud (See \S \ref{sec:morp}). Second, most SMUDGes galaxies with a redshift measurement are located in relatively dense environments (i.e., galaxy groups or clusters), which preferentially host redder SMUDGes \citep{kadowaki21}. 
\cite{neumayer} highlight similar biases in previous studies and concluded that there is yet no evidence for a strong morphological distinction in the occupation fractions. 

We conclude that the morphological dependence of $f_{NSC}$ is potentially a constraining factor for scenarios but firmer theoretical predictions are needed to establish whether this is indeed a discriminating factor and further observational work, addressing observational biases, is needed to establish any empirical results along these lines.

\subsection{Are There Isolated NSCs?}

A motivation for our SDSS work was to explore the possibility that there exist galaxies of even lower central surface brightness than those cataloged in SMUDGes that host an NSC. Such objects would appear as un-hosted NSCs. 
It is possible to discover galaxies through their dense collections of stars, such as GCs or NSCs \citep{zaritsky16}.
Such systems, if found, would appear to be difficult to model as the result of a major merger that resulted in significant in-situ star formation and hence an interesting constraint on NSC formation models.

We identified 18 SDSS unresolved sources that have a large recessional velocity (cz $>$ 1000 km s$^{-1}$) as described in Figure \ref{fig:bare_psf} and \S \ref{sec:bare_psf}. The bluer objects in this sample could be hyper-velocity white dwarfs that are thought to be runaways from Type Ia supernovae \citep{2023OJAp....6E..28E} although they could also arise from spurious velocity measurements, and the redder objects mostly lie within the Virgo cluster and/or nearby giant galaxies, suggesting an interaction-specific evolutionary process. 
The identified Bare PSFs near the Virgo cluster (Objects 6, 8, 9, 10, 11, 12, 13, 15, 16, and 17 in Figure \ref{fig:bare_psf}) have a median $r$-band absolute magnitude of $-10.56\pm0.68$ mag, and are therefore much brighter than the typical globular cluster \citep[absolute magnitude approximately $-7$ mag, e.g.,][]{2009Jordan_GC_NSC_Lum}. The median stellar mass of those objects is $10^{6.74\pm0.02}$\Msol. These objects appear unresolved in the image at the distance of the Virgo cluster, indicating that their physical effective radii are smaller than 110 pc. Given their compact size and luminosity, these objects could likely be classified as ultra-compact dwarfs \citep[UCDs;][]{Hilker1999, Drinkwater2000}. Recently, \cite{Wang+23} found 106 galaxies within the Virgo cluster having morphologies that bridge the gap between nucleated dwarf galaxies and UCDs, indicating a transitional stage from NSCs to UCDs. Their research aligns with our own findings.

UCDs are proposed to be NSCs where the host galaxy has been stripped away \citep{Drinkwater2003, Pfeffer2013, Norris2014, Dumont2022}. 
Object 15 in Figure \ref{fig:bare_psf} appears to have an anisotropic tail and so might be in a transitional stage from NSC-bearing galaxy to UCD. Alternatively, at least some of these objects may still retain a faint underlying S\'ersic profile that eluded detection given their higher-than-typical local backgrounds.

We lack a definitive explanation for the single isolated red bare PSF that is not spectroscopically classified as a white dwarf. 
This object could be a stripped NSC that is not obviously in a dense environment, it might also have an extremely faint underlying S\'ersic profile that eludes detection, it might be either a mis-classified white dwarf or a different type of hypervelocity star, or, finally, it could just be a spurious velocity measurement. Regardless, while interesting, such objects are clearly a small fraction of the overall sample. The exclusion of this object, should it actually have a faint host, does not affect our results nor enable robust discrimination between NSC formation scenarios.

% \begin{figure}
% 	\includegraphics[width=\columnwidth]{images/paper_smdgdist_nsc2dfrac_AREnv.png}
%     \caption{The dependence of NSC occupation fraction, $f_{NSC}$, on axis ratio and the environmental measure D$_{10}$ for the SMUDGes sample.}
%     \label{fig:paper_smdgdist_nsc2dfrac_AREnv}
% \end{figure}

\begin{figure}
	\includegraphics[width=\columnwidth]{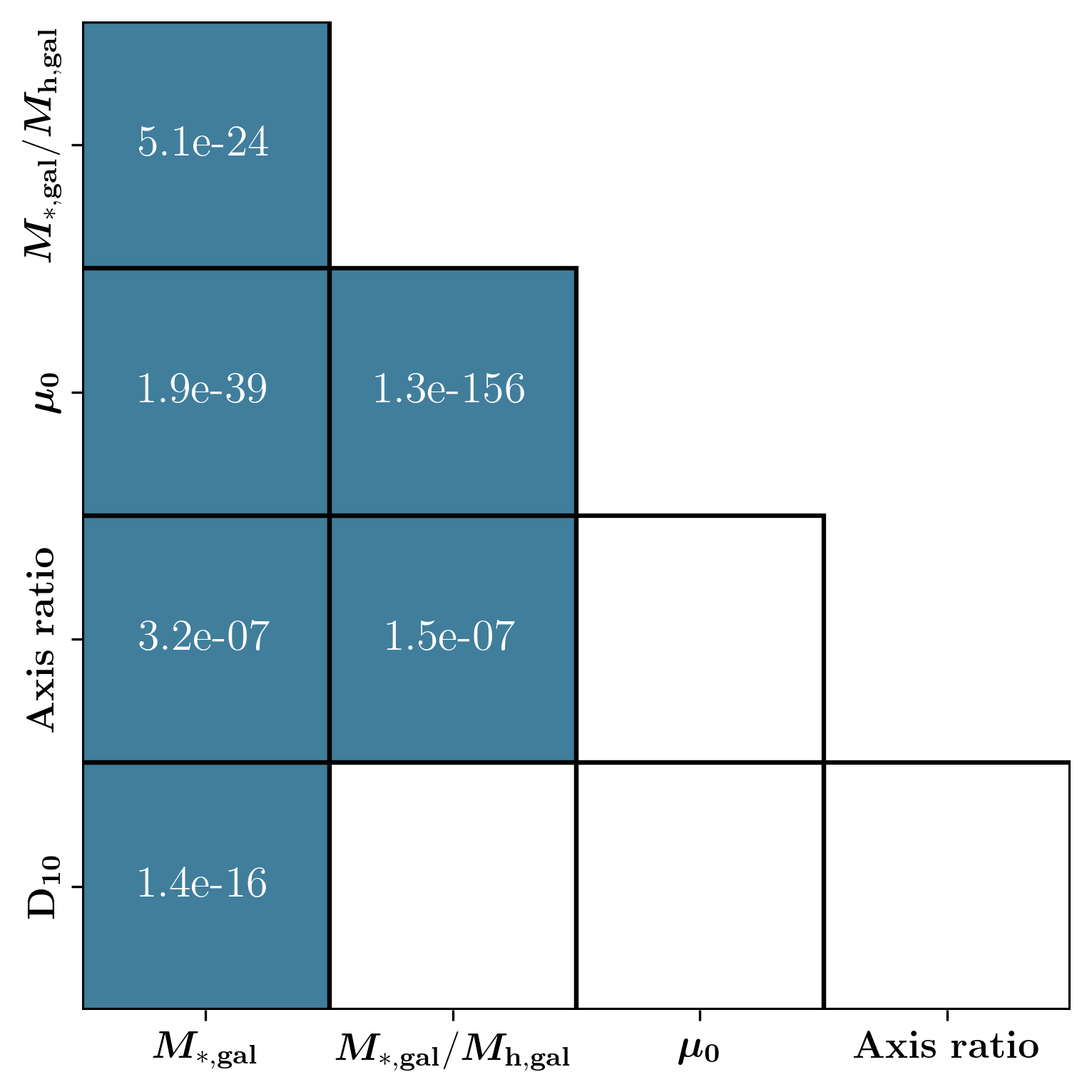}
    \caption{The correlation matrix displays the parameters analyzed for the NSC occupation fraction. The values within each cell represent the p-values obtained from the Spearman rank correlation test. We display only the cells where the p-value is less than 0.01.}
    \label{fig:corr_matrix}
\end{figure}

\subsection{What Drives the Occupation Fraction?}

We find that the NSC occupation fraction ($f_{NSC}$) correlates with  1) stellar mass, 2) stellar mass fraction, 3) central surface brightness, 4) host galaxy axis ratio, and 5) environment (D$_{10}$).
The difficulty in interpreting these trends is that the different properties are not independent.

We examine the correlation matrix, using Spearman rank correlations, for these properties (Figure \ref{fig:corr_matrix}) 
to assess whether any one or two of these could account for all of the observed behavior.
We find that $M_{*,gal}$ correlates with all of the other quantities we have considered. As such, if $f_{NSC}$ is solely driven by $M_{*,gal}$, we would still expect to observe correlations between $f_{NSC}$ and the other quantities. On the other hand, we concluded from Figure \ref{fig:smdgdist_nsc2dfrac_MhM} that stellar mass fraction ($M_{*,gal}/M_{h,gal}$) has a role in determining $f_{NSC}$ that is independent of $M_{*,gal}$. Positing that stellar mass fraction is instead a driver of $f_{NSC}$, the lack of a correlation between stellar mass fraction and D$_{10}$ implies that environment plays a role as well in determining $f_{NSC}$. 

While this type of analysis does not allow us to definitively identify only one or two drivers of $f_{NSC}$, it does highlight that finding that many factors correlate with $f_{NSC}$ does imply that they all play a physical role. At least in broad strokes, one can speculate that a limited number of these factors do indeed play such a role. One can imagine carrying out a similar quantitative analysis, perhaps examining residual correlations. However, the challenge lies in the qualitative differences in the measurement uncertainties. For example, our measurement of the axis ratio is entirely empirical, while our measurement of the stellar mass fraction is almost certainly dominated by systematic uncertainties in our estimates of $M_{*,gal}$ and $M_{h,gal}$.

\subsection{The $M_{NSC}$-$M_{h,gal}$ relation}

One of our principal results, the direct proportionality between $M_{NSC}$ with $M_{h,gal}$ depends critically on the highly uncertain estimation of $M_{h,gal}$, but also offers tremendous potential for discriminating among formation models. To be clear, we see no reason that in-situ formation of an NSC would lead to such a relation. As such, it is imperative to understand if the finding is correct or simply ``fortuitous".

Alternative estimates of $M_{h,gal}$, for a different sample of nucleated galaxies, have yielded a different conclusion \citep{sanchez2019}. In that study, abundance matching was used to transform $M_*$ into $M_{h,gal}$ \citep{grossauer}, which resulted in a superlinear relation between $M_{NSC}$ and $M_{h,gal}$ ($M_{NSC} \propto M_{h,gal}^{1.2}$). Those authors did note the uncertain nature of the $M_{h,gal}$ estimates. The difference between $M_{h,gal}$ estimates from \cite{grossauer} and the method we adopted here was already noted by \cite{2023MNRAS.519..871Z}. Establishing which, if either, approach yields accurate masses awaits direct measurements of the mass of these low-mass galaxies. We do, however, argue that the resulting linear relations we find between $M_{NSC}$ and $M_{h,gal}$, and that found previously between N$_{GC}$ and $M_{h,gal}$ \citep{2022Zaritsky_GC} using the same mass estimation method we adopted here, suggest that the adopted method for estimating $M_{h,gal}$ is yielding accurate results.

\subsection{The GCs and NSCs Mass Fractions}

One simple constraint on NSC formation models is the mass fraction in NSCs vs. GCs. For example, if the NSC mass fraction far exceeds the GC mass fraction, then NSCs could manifestly not be created from inspiraling GCs. 

From our fitting to the $M_{NSC}-M_{h,gal}$ relation we find a mass fraction of $\sim$ $10^{-4}$ for NSCs. For comparison, the mass fraction in globular clusters is estimated to be $\sim 5\times 10^{-5}$ \citep{forbes}, and while smaller than the NSC mass fraction it is so by only a factor of two. With uncertainties in how the halo and stellar masses are estimated between our study and that of \cite{forbes}, these fractions are effectively in agreement. Other studies reached consistent conclusions 
\citep{Cote2006,denBrok2014}. A uniform analysis, using the same photometric data and methodology for galaxies with and without NSC and their GC systems, will address whether the mass fractions are indeed consistent or differ by as much as about a factor of two.

The near equality of these mass fractions, as well as the large fraction of systems without an NSC, would appear to argue 
that in an infalling GC scenario either most or none of the GCs find their way to the center. If this is borne out by more precise measurements of $M_{NSC}$ and GC population studies, it may prove to be a critical piece of information. 
A simple test, assuming that the NSC and GC fractions are nearly equal, would be to confirm that the NSC-bearing galaxies are relatively devoid of GCs.
On the other hand, the near equality in mass fractions appears difficult to reconcile with an in-situ star formation model, where one could imagine forming central clusters with a mass that is independent of $M_{h,gal}$.

\begin{figure}
	\includegraphics[width=\columnwidth]{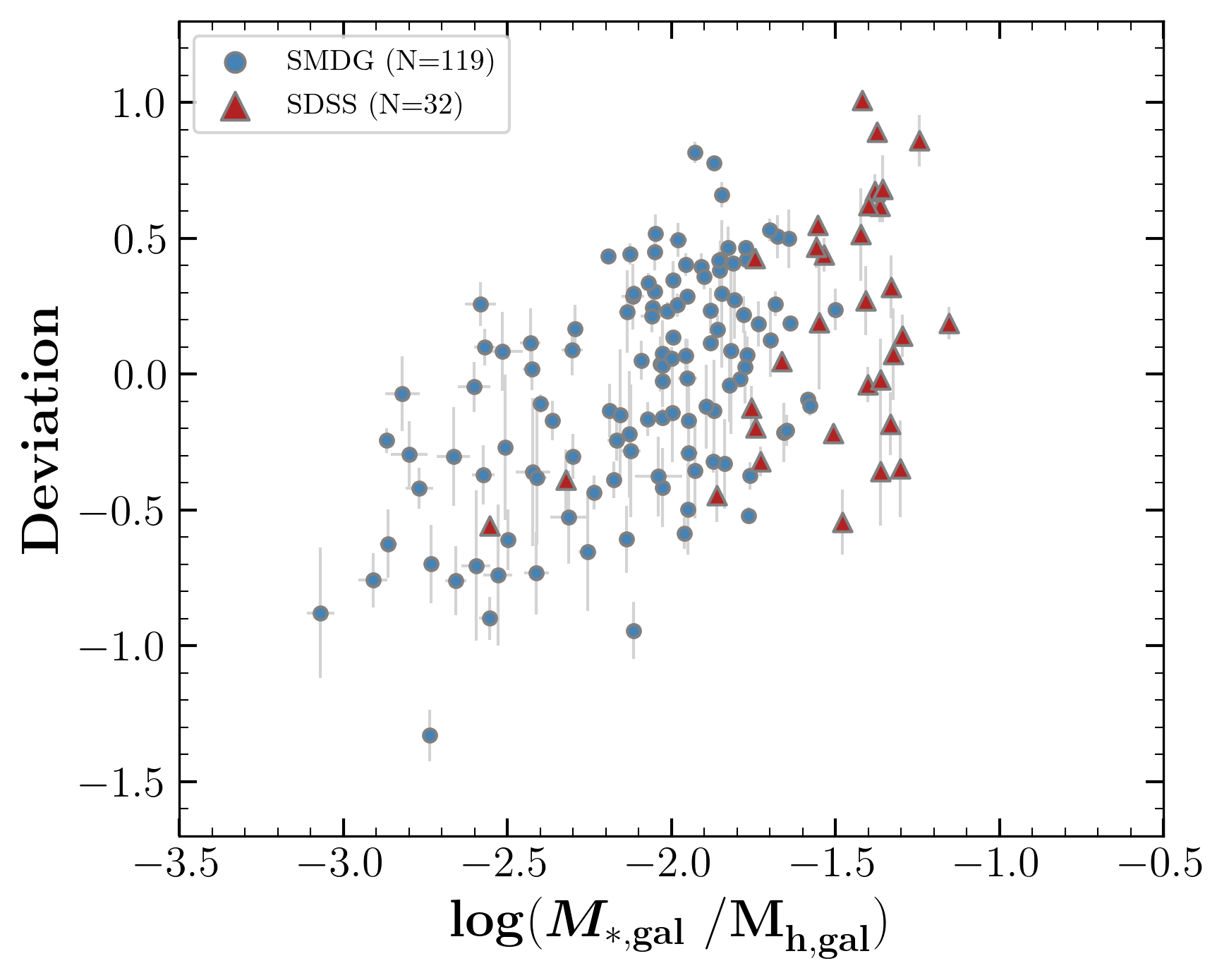}
    \caption{The deviation from the $M_{NSC}$-$M_{h,gal}$ relation (Figure \ref{fig:both_mass_hmass}) and the stellar mass fraction. Galaxies above the best-fit line in Figure \ref{fig:both_mass_hmass} tend to have a higher stellar mass fraction. The p-value from the Spearman rank correlation test is 3.28e-10.}
    \label{fig:dev_mfrac}
\end{figure}

The situation, however, may be more complex. 
While Figure \ref{fig:both_mass_hmass} shows a direct proportional linear relation between $M_{NSC}$ and $M_{h,gal}$, the deviation from the best-fit line is correlated with the stellar mass fraction, $M_{*,gal}/M_{h,gal}$, of the host galaxy (Figure \ref{fig:dev_mfrac}). One possibility is that infalling GCs play the principal role in the formation of NSCs and in-situ star formation makes a secondary contribution. Some numerical simulations support the idea that both formation mechanisms are required to reproduce the mass of NSCs
\citep{Hartmann2011, Antonini2012}. On the other hand, it is also possible that this relation is an artifact created by how our $M_{h,gal}$ estimator inadequately treats the baryonic contribution.

\section{Summary}
\label{sec:summary}

We analyze NSCs in galaxies using two datasets: SMUDGes for low surface brightness galaxies ($\langle \mu_{0,g}\rangle = 24.8$ mag arcsec$^{-2}$), and SDSS for relatively brighter galaxies ($\langle \mu_{0,g}\rangle = 21.2$ mag arcsec$^{-2}$). With $g$- and $r$-band images from the Legacy Survey DR9 database, we develop a two-stage fitting approach that involves fitting single S\'ersic profiles and simultaneous decomposition into multiple components. 
We identify a total of 578 and 42 targets in SMUDGes and SDSS, respectively, with unresolved point sources at more than 90\% confidence. Among these, 273 and 32 targets, respectively, have point sources that are less than 0.1 $r_e$ offset from the center of their host. Previous work \citep{lambert} found that among such NSC candidates there is a modest contamination fraction (0.15).
The investigation into detection limits using simulations indicates that the 50\% completeness magnitude for the SMUDGes sample is 24.1 mag, while that for the SDSS sample is 23.4 mag. This analysis implies a higher likelihood of undetected NSCs in the SDSS sample.

To investigate the relationship between NSC properties and their host galaxies, we derive estimates for the stellar and total host galaxy masses. The stellar mass ($M_{*,gal}$) is determined by summing the contributions of all S\'ersic components, excluding the NSC. The NSC stellar mass ($M_{NSC}$) is calculated using a similar method, but uncertainties are higher due to sensitivity to PSF uncertainties. The halo mass ($M_{h,gal}$) is estimated with a photometric halo mass estimator proposed by \cite{2023MNRAS.519..871Z}. We also measure D$_{10}$, the distance to the 10th nearest galaxy, as a measure of the galactic environment. D$_{10}$ is derived from a set of magnitude-limited SDSS galaxies within the SDSS footprint.

We first examine the morphological distribution of NSC-bearing galaxies, using $g-r$ as a proxy for morphology. Most of our NSC hosts lie on the red sequence. The absence of blue dwarfs in SMUDGes and a notable proportion of `failed fitting' galaxies in SDSS could result in an undercounting of NSCs in blue galaxies. Thus, our following conclusions are predominantly relevant to red galaxies, and addressing questions about NSC properties across various host galaxy morphologies proves challenging with our sample.

One of our main results is that the fraction of low surface brightness galaxies hosting NSCs varies strongly as a function of a set of galaxy characteristics. We find an increase in this occupation fraction when: 1) the stellar mass increases, 2) the stellar mass fraction ($M_{*,gal}/M_{h,gal}$) increases, 3) the central surface brightness becomes brighter, 4) the host galaxy axis ratio increases or 5) the environment becomes denser (as indicated by a decrease in D$_{10}$). 
From our analysis of the correlation matrix among these quantities, we conclude that a combination of the galaxy stellar mass fraction and environment could serve as the drivers of the occupation fraction. Nevertheless, it is essential to clarify that this is not the only possibility and that further work must be done to untangle this set of correlations.

We also investigate the scaling relations for the mass of the NSC ($M_{NSC}$). The NSC mass correlates with the stellar mass of its host galaxy. Using the bisector ordinary least squares fitting method we find that $M_{NSC}/$\Msol$ = 10^{6.02\pm0.03}(M_{*,gal}/10^{8} $\Msol$)^{0.77\pm0.04}$. 
A comparison with the result obtained using the traditional ordinary least squares method highlights significant differences in the derived slopes ($M_{NSC}/$\Msol$ = 10^{5.95\pm0.04}(M_{*,gal}/10^{8} $\Msol$)^{0.64\pm0.04}$), emphasizing the sensitivity of results to fitting methodology. The NSC mass also correlates with the halo mass of its host ($M_{NSC}/$\Msol$ = 10^{6.11\pm0.05}(M_{h,gal}/10^{10} $\Msol$)^{0.92\pm0.05}$). Notably, the slope of the fit suggests a direct proportionality between $M_{NSC}$ and $M_{h,gal}$ and that the fraction of the total mass that is in an NSC, in galaxies with an NSC, is $\sim 10^{-4}$. The proportionality of the NSC mass to the halo mass and the value of the mass fraction, both argue for a close relationship between GC and NSC formation.

\begin{acknowledgments}
The authors acknowledge financial support from NSF AST-1713841 and AST-2006785.  An allocation of computer time from the UA Research Computing High Performance Computing (HPC) at the University of Arizona and the prompt assistance of the associated computer support group is gratefully acknowledged.

Funding for the Sloan Digital Sky Survey IV has been provided by the Alfred P. Sloan Foundation, the U.S. Department of Energy Office of Science, and the Participating Institutions. SDSS acknowledges support and resources from the Center for High-Performance Computing at the University of Utah. The SDSS web site is www.sdss4.org. This research depends directly on images from the Dark Energy Camera Legacy Survey (DECaLS; Proposal ID 2014B-0404; PIs: David Schlegel and Arjun Dey). Full acknowledgment at https://www.legacysurvey.org/acknowledgment/.

\end{acknowledgments}
\software{
\texttt{Astropy }             \citep{astropy1, astropy2},
\texttt{astroquery  }         \citep{astroquery},
\texttt{GALFIT }              \citep{peng},
\texttt{Matplotlib }          \citep{matplotlib},
\texttt{NumPy }               \citep{numpy},
\texttt{pandas }              \citep{pandas},
\texttt{SEP}                \citep{sep},
\texttt{Source Extractor}     \citep{bertin},
\texttt{SciPy}                \citep{scipy1, scipy2},
\texttt{dustmaps}   \citep{green}
}

\bibliography{references.bib}{}

\begin{thebibliography}{}
\expandafter\ifx\csname natexlab\endcsname\relax\def\natexlab#1{#1}\fi
\providecommand{\url}[1]{\href{#1}{#1}}
\providecommand{\dodoi}[1]{doi:~\href{http://doi.org/#1}{\nolinkurl{#1}}}
\providecommand{\doeprint}[1]{\href{http://ascl.net/#1}{\nolinkurl{http://ascl.net/#1}}}
\providecommand{\doarXiv}[1]{\href{https://arxiv.org/abs/#1}{\nolinkurl{https://arxiv.org/abs/#1}}}

\bibitem[{{Ahumada} {et~al.}(2020){Ahumada}, {Allende Prieto}, {Almeida}, {Anders}, {Anderson}, {Andrews}, {Anguiano}, {Arcodia}, {Armengaud}, {Aubert}, {Avila}, {Avila-Reese}, {Badenes}, {Balland}, {Barger}, {Barrera-Ballesteros}, {Basu}, {Bautista}, {Beaton}, {Beers}, {Benavides}, {Bender}, {Bernardi}, {Bershady}, {Beutler}, {Bidin}, {Bird}, {Bizyaev}, {Blanc}, {Blanton}, {Boquien}, {Borissova}, {Bovy}, {Brandt}, {Brinkmann}, {Brownstein}, {Bundy}, {Bureau}, {Burgasser}, {Burtin}, {Cano-D{\'\i}az}, {Capasso}, {Cappellari}, {Carrera}, {Chabanier}, {Chaplin}, {Chapman}, {Cherinka}, {Chiappini}, {Doohyun Choi}, {Chojnowski}, {Chung}, {Clerc}, {Coffey}, {Comerford}, {Comparat}, {da Costa}, {Cousinou}, {Covey}, {Crane}, {Cunha}, {Ilha}, {Dai}, {Damsted}, {Darling}, {Davidson}, {Davies}, {Dawson}, {De}, {de la Macorra}, {De Lee}, {Queiroz}, {Deconto Machado}, {de la Torre}, {Dell'Agli}, {du Mas des Bourboux}, {Diamond-Stanic}, {Dillon}, {Donor}, {Drory}, {Duckworth}, {Dwelly}, {Ebelke}, {Eftekharzadeh}, {Davis
  Eigenbrot}, {Elsworth}, {Eracleous}, {Erfanianfar}, {Escoffier}, {Fan}, {Farr}, {Fern{\'a}ndez-Trincado}, {Feuillet}, {Finoguenov}, {Fofie}, {Fraser-McKelvie}, {Frinchaboy}, {Fromenteau}, {Fu}, {Galbany}, {Garcia}, {Garc{\'\i}a-Hern{\'a}ndez}, {Garma Oehmichen}, {Ge}, {Geimba Maia}, {Geisler}, {Gelfand}, {Goddy}, {Gonzalez-Perez}, {Grabowski}, {Green}, {Grier}, {Guo}, {Guy}, {Harding}, {Hasselquist}, {Hawken}, {Hayes}, {Hearty}, {Hekker}, {Hogg}, {Holtzman}, {Horta}, {Hou}, {Hsieh}, {Huber}, {Hunt}, {Ider Chitham}, {Imig}, {Jaber}, {Jimenez Angel}, {Johnson}, {Jones}, {J{\"o}nsson}, {Jullo}, {Kim}, {Kinemuchi}, {Kirkpatrick}, {Kite}, {Klaene}, {Kneib}, {Kollmeier}, {Kong}, {Kounkel}, {Krishnarao}, {Lacerna}, {Lan}, {Lane}, {Law}, {Le Goff}, {Leung}, {Lewis}, {Li}, {Lian}, {Lin}, {Long}, {Longa-Pe{\~n}a}, {Lundgren}, {Lyke}, {Mackereth}, {MacLeod}, {Majewski}, {Manchado}, {Maraston}, {Martini}, {Masseron}, {Masters}, {Mathur}, {McDermid}, {Merloni}, {Merrifield}, {M{\'e}sz{\'a}ros}, {Miglio}, {Minniti},
  {Minsley}, {Miyaji}, {Mohammad}, {Mosser}, {Mueller}, {Muna}, {Mu{\~n}oz-Guti{\'e}rrez}, {Myers}, {Nadathur}, {Nair}, {Nandra}, {Correa do Nascimento}, {Nevin}, {Newman}, {Nidever}, {Nitschelm}, {Noterdaeme}, {O'Connell}, {Olmstead}, {Oravetz}, {Oravetz}, {Osorio}, {Pace}, {Padilla}, {Palanque-Delabrouille}, {Palicio}, {Pan}, {Pan}, {Parker}, {Paviot}, {Peirani}, {Ram{\'r}ez}, {Penny}, {Percival}, {Perez-Fournon}, {P{\'e}rez-R{\`a}fols}, {Petitjean}, {Pieri}, {Pinsonneault}, {Poovelil}, {Povick}, {Prakash}, {Price-Whelan}, {Raddick}, {Raichoor}, {Ray}, {Rembold}, {Rezaie}, {Riffel}, {Riffel}, {Rix}, {Robin}, {Roman-Lopes}, {Rom{\'a}n-Z{\'u}{\~n}iga}, {Rose}, {Ross}, {Rossi}, {Rowlands}, {Rubin}, {Salvato}, {S{\'a}nchez}, {S{\'a}nchez-Menguiano}, {S{\'a}nchez-Gallego}, {Sayres}, {Schaefer}, {Schiavon}, {Schimoia}, {Schlafly}, {Schlegel}, {Schneider}, {Schultheis}, {Schwope}, {Seo}, {Serenelli}, {Shafieloo}, {Shamsi}, {Shao}, {Shen}, {Shetrone}, {Shirley}, {Silva Aguirre}, {Simon}, {Skrutskie}, {Slosar},
  {Smethurst}, {Sobeck}, {Sodi}, {Souto}, {Stark}, {Stassun}, {Steinmetz}, {Stello}, {Stermer}, {Storchi-Bergmann}, {Streblyanska}, {Stringfellow}, {Stutz}, {Su{\'a}rez}, {Sun}, {Taghizadeh-Popp}, {Talbot}, {Tayar}, {Thakar}, {Theriault}, {Thomas}, {Thomas}, {Tinker}, {Tojeiro}, {Toledo}, {Tremonti}, {Troup}, {Tuttle}, {Unda-Sanzana}, {Valentini}, {Vargas-Gonz{\'a}lez}, {Vargas-Maga{\~n}a}, {V{\'a}zquez-Mata}, {Vivek}, {Wake}, {Wang}, {Weaver}, {Weijmans}, {Wild}, {Wilson}, {Wilson}, {Wolthuis}, {Wood-Vasey}, {Yan}, {Yang}, {Y{\`e}che}, {Zamora}, {Zarrouk}, {Zasowski}, {Zhang}, {Zhao}, {Zhao}, {Zheng}, {Zheng}, {Zhu}, \& {Zou}}]{2020ApJS..249....3A}
{Ahumada}, R., {Allende Prieto}, C., {Almeida}, A., {et~al.} 2020, \apjs, 249, 3, \dodoi{10.3847/1538-4365/ab929e}

\bibitem[{Akaike(1974)}]{AIC}
Akaike, H. 1974, IEEE Transactions on Automatic Control, 19, 716, \dodoi{10.1109/TAC.1974.1100705}

\bibitem[{{Antonini} {et~al.}(2012){Antonini}, {Capuzzo-Dolcetta}, {Mastrobuono-Battisti}, \& {Merritt}}]{Antonini2012}
{Antonini}, F., {Capuzzo-Dolcetta}, R., {Mastrobuono-Battisti}, A., \& {Merritt}, D. 2012, \apj, 750, 111, \dodoi{10.1088/0004-637X/750/2/111}

\bibitem[{{Astropy Collaboration} {et~al.}(2013){Astropy Collaboration}, {Robitaille}, {Tollerud}, {Greenfield}, {Droettboom}, {Bray}, {Aldcroft}, {Davis}, {Ginsburg}, {Price-Whelan}, {Kerzendorf}, {Conley}, {Crighton}, {Barbary}, {Muna}, {Ferguson}, {Grollier}, {Parikh}, {Nair}, {Unther}, {Deil}, {Woillez}, {Conseil}, {Kramer}, {Turner}, {Singer}, {Fox}, {Weaver}, {Zabalza}, {Edwards}, {Azalee Bostroem}, {Burke}, {Casey}, {Crawford}, {Dencheva}, {Ely}, {Jenness}, {Labrie}, {Lim}, {Pierfederici}, {Pontzen}, {Ptak}, {Refsdal}, {Servillat}, \& {Streicher}}]{astropy1}
{Astropy Collaboration}, {Robitaille}, T.~P., {Tollerud}, E.~J., {et~al.} 2013, \aap, 558, A33, \dodoi{10.1051/0004-6361/201322068}

\bibitem[{{Astropy Collaboration} {et~al.}(2018){Astropy Collaboration}, {Price-Whelan}, {Sip{\H{o}}cz}, {G{\"u}nther}, {Lim}, {Crawford}, {Conseil}, {Shupe}, {Craig}, {Dencheva}, {Ginsburg}, {VanderPlas}, {Bradley}, {P{\'e}rez-Su{\'a}rez}, {de Val-Borro}, {Aldcroft}, {Cruz}, {Robitaille}, {Tollerud}, {Ardelean}, {Babej}, {Bach}, {Bachetti}, {Bakanov}, {Bamford}, {Barentsen}, {Barmby}, {Baumbach}, {Berry}, {Biscani}, {Boquien}, {Bostroem}, {Bouma}, {Brammer}, {Bray}, {Breytenbach}, {Buddelmeijer}, {Burke}, {Calderone}, {Cano Rodr{\'\i}guez}, {Cara}, {Cardoso}, {Cheedella}, {Copin}, {Corrales}, {Crichton}, {D'Avella}, {Deil}, {Depagne}, {Dietrich}, {Donath}, {Droettboom}, {Earl}, {Erben}, {Fabbro}, {Ferreira}, {Finethy}, {Fox}, {Garrison}, {Gibbons}, {Goldstein}, {Gommers}, {Greco}, {Greenfield}, {Groener}, {Grollier}, {Hagen}, {Hirst}, {Homeier}, {Horton}, {Hosseinzadeh}, {Hu}, {Hunkeler}, {Ivezi{\'c}}, {Jain}, {Jenness}, {Kanarek}, {Kendrew}, {Kern}, {Kerzendorf}, {Khvalko}, {King}, {Kirkby}, {Kulkarni},
  {Kumar}, {Lee}, {Lenz}, {Littlefair}, {Ma}, {Macleod}, {Mastropietro}, {McCully}, {Montagnac}, {Morris}, {Mueller}, {Mumford}, {Muna}, {Murphy}, {Nelson}, {Nguyen}, {Ninan}, {N{\"o}the}, {Ogaz}, {Oh}, {Parejko}, {Parley}, {Pascual}, {Patil}, {Patil}, {Plunkett}, {Prochaska}, {Rastogi}, {Reddy Janga}, {Sabater}, {Sakurikar}, {Seifert}, {Sherbert}, {Sherwood-Taylor}, {Shih}, {Sick}, {Silbiger}, {Singanamalla}, {Singer}, {Sladen}, {Sooley}, {Sornarajah}, {Streicher}, {Teuben}, {Thomas}, {Tremblay}, {Turner}, {Terr{\'o}n}, {van Kerkwijk}, {de la Vega}, {Watkins}, {Weaver}, {Whitmore}, {Woillez}, {Zabalza}, \& {Astropy Contributors}}]{astropy2}
{Astropy Collaboration}, {Price-Whelan}, A.~M., {Sip{\H{o}}cz}, B.~M., {et~al.} 2018, \aj, 156, 123, \dodoi{10.3847/1538-3881/aabc4f}

\bibitem[{{Bailey}(1980)}]{Bailey1980}
{Bailey}, M.~E. 1980, \mnras, 191, 195, \dodoi{10.1093/mnras/191.2.195}

\bibitem[{{Baldassare} {et~al.}(2014){Baldassare}, {Gallo}, {Miller}, {Plotkin}, {Treu}, {Valluri}, \& {Woo}}]{Baldassare2014}
{Baldassare}, V.~F., {Gallo}, E., {Miller}, B.~P., {et~al.} 2014, \apj, 791, 133, \dodoi{10.1088/0004-637X/791/2/133}

\bibitem[{Barbary(2016)}]{sep}
Barbary, K. 2016, {SEP: Source Extractor as a library}, \dodoi{10.21105/joss.00058}

\bibitem[{{Beasley} {et~al.}(2016){Beasley}, {Romanowsky}, {Pota}, {Navarro}, {Martinez Delgado}, {Neyer}, \& {Deich}}]{Beasley2016}
{Beasley}, M.~A., {Romanowsky}, A.~J., {Pota}, V., {et~al.} 2016, \apjl, 819, L20, \dodoi{10.3847/2041-8205/819/2/L20}

\bibitem[{{Bekki} \& {Couch}(2001)}]{bekki01}
{Bekki}, K., \& {Couch}, W.~J. 2001, \apjl, 557, L19, \dodoi{10.1086/323139}

\bibitem[{{Bertin} \& {Arnouts}(1996)}]{bertin}
{Bertin}, E., \& {Arnouts}, S. 1996, \aaps, 117, 393, \dodoi{10.1051/aas:1996164}

\bibitem[{{Binggeli} {et~al.}(1987){Binggeli}, {Tammann}, \& {Sandage}}]{binggeli}
{Binggeli}, B., {Tammann}, G.~A., \& {Sandage}, A. 1987, \aj, 94, 251, \dodoi{10.1086/114467}

\bibitem[{{Blakeslee} {et~al.}(1997){Blakeslee}, {Tonry}, \& {Metzger}}]{blakeslee}
{Blakeslee}, J.~P., {Tonry}, J.~L., \& {Metzger}, M.~R. 1997, \aj, 114, 482, \dodoi{10.1086/118488}

\bibitem[{{Blanton} {et~al.}(2017){Blanton}, {Bershady}, {Abolfathi}, {Albareti}, {Allende Prieto}, {Almeida}, {Alonso-Garc{\'\i}a}, {Anders}, {Anderson}, {Andrews}, {Aquino-Ort{\'\i}z}, {Arag{\'o}n-Salamanca}, {Argudo-Fern{\'a}ndez}, {Armengaud}, {Aubourg}, {Avila-Reese}, {Badenes}, {Bailey}, {Barger}, {Barrera-Ballesteros}, {Bartosz}, {Bates}, {Baumgarten}, {Bautista}, {Beaton}, {Beers}, {Belfiore}, {Bender}, {Berlind}, {Bernardi}, {Beutler}, {Bird}, {Bizyaev}, {Blanc}, {Blomqvist}, {Bolton}, {Boquien}, {Borissova}, {van den Bosch}, {Bovy}, {Brandt}, {Brinkmann}, {Brownstein}, {Bundy}, {Burgasser}, {Burtin}, {Busca}, {Cappellari}, {Delgado Carigi}, {Carlberg}, {Carnero Rosell}, {Carrera}, {Chanover}, {Cherinka}, {Cheung}, {G{\'o}mez Maqueo Chew}, {Chiappini}, {Choi}, {Chojnowski}, {Chuang}, {Chung}, {Cirolini}, {Clerc}, {Cohen}, {Comparat}, {da Costa}, {Cousinou}, {Covey}, {Crane}, {Croft}, {Cruz-Gonzalez}, {Garrido Cuadra}, {Cunha}, {Damke}, {Darling}, {Davies}, {Dawson}, {de la Macorra}, {Dell'Agli}, {De
  Lee}, {Delubac}, {Di Mille}, {Diamond-Stanic}, {Cano-D{\'\i}az}, {Donor}, {Downes}, {Drory}, {du Mas des Bourboux}, {Duckworth}, {Dwelly}, {Dyer}, {Ebelke}, {Eigenbrot}, {Eisenstein}, {Emsellem}, {Eracleous}, {Escoffier}, {Evans}, {Fan}, {Fern{\'a}ndez-Alvar}, {Fernandez-Trincado}, {Feuillet}, {Finoguenov}, {Fleming}, {Font-Ribera}, {Fredrickson}, {Freischlad}, {Frinchaboy}, {Fuentes}, {Galbany}, {Garcia-Dias}, {Garc{\'\i}a-Hern{\'a}ndez}, {Gaulme}, {Geisler}, {Gelfand}, {Gil-Mar{\'\i}n}, {Gillespie}, {Goddard}, {Gonzalez-Perez}, {Grabowski}, {Green}, {Grier}, {Gunn}, {Guo}, {Guy}, {Hagen}, {Hahn}, {Hall}, {Harding}, {Hasselquist}, {Hawley}, {Hearty}, {Gonzalez Hern{\'a}ndez}, {Ho}, {Hogg}, {Holley-Bockelmann}, {Holtzman}, {Holzer}, {Huehnerhoff}, {Hutchinson}, {Hwang}, {Ibarra-Medel}, {da Silva Ilha}, {Ivans}, {Ivory}, {Jackson}, {Jensen}, {Johnson}, {Jones}, {J{\"o}nsson}, {Jullo}, {Kamble}, {Kinemuchi}, {Kirkby}, {Kitaura}, {Klaene}, {Knapp}, {Kneib}, {Kollmeier}, {Lacerna}, {Lane}, {Lang}, {Law},
  {Lazarz}, {Lee}, {Le Goff}, {Liang}, {Li}, {Li}, {Lian}, {Lima}, {Lin}, {Lin}, {Bertran de Lis}, {Liu}, {de Icaza Lizaola}, {Long}, {Lucatello}, {Lundgren}, {MacDonald}, {Deconto Machado}, {MacLeod}, {Mahadevan}, {Geimba Maia}, {Maiolino}, {Majewski}, {Malanushenko}, {Malanushenko}, {Manchado}, {Mao}, {Maraston}, {Marques-Chaves}, {Masseron}, {Masters}, {McBride}, {McDermid}, {McGrath}, {McGreer}, {Medina Pe{\~n}a}, {Melendez}, {Merloni}, {Merrifield}, {Meszaros}, {Meza}, {Minchev}, {Minniti}, {Miyaji}, {More}, {Mulchaey}, {M{\"u}ller-S{\'a}nchez}, {Muna}, {Munoz}, {Myers}, {Nair}, {Nandra}, {Correa do Nascimento}, {Negrete}, {Ness}, {Newman}, {Nichol}, {Nidever}, {Nitschelm}, {Ntelis}, {O'Connell}, {Oelkers}, {Oravetz}, {Oravetz}, {Pace}, {Padilla}, {Palanque-Delabrouille}, {Alonso Palicio}, {Pan}, {Parejko}, {Parikh}, {P{\^a}ris}, {Park}, {Patten}, {Peirani}, {Pellejero-Ibanez}, {Penny}, {Percival}, {Perez-Fournon}, {Petitjean}, {Pieri}, {Pinsonneault}, {Pisani}, {Poleski}, {Prada}, {Prakash}, {Queiroz},
  {Raddick}, {Raichoor}, {Barboza Rembold}, {Richstein}, {Riffel}, {Riffel}, {Rix}, {Robin}, {Rockosi}, {Rodr{\'\i}guez-Torres}, {Roman-Lopes}, {Rom{\'a}n-Z{\'u}{\~n}iga}, {Rosado}, {Ross}, {Rossi}, {Ruan}, {Ruggeri}, {Rykoff}, {Salazar-Albornoz}, {Salvato}, {S{\'a}nchez}, {Aguado}, {S{\'a}nchez-Gallego}, {Santana}, {Santiago}, {Sayres}, {Schiavon}, {da Silva Schimoia}, {Schlafly}, {Schlegel}, {Schneider}, {Schultheis}, {Schuster}, {Schwope}, {Seo}, {Shao}, {Shen}, {Shetrone}, {Shull}, {Simon}, {Skinner}, {Skrutskie}, {Slosar}, {Smith}, {Sobeck}, {Sobreira}, {Somers}, {Souto}, {Stark}, {Stassun}, {Stauffer}, {Steinmetz}, {Storchi-Bergmann}, {Streblyanska}, {Stringfellow}, {Su{\'a}rez}, {Sun}, {Suzuki}, {Szigeti}, {Taghizadeh-Popp}, {Tang}, {Tao}, {Tayar}, {Tembe}, {Teske}, {Thakar}, {Thomas}, {Thompson}, {Tinker}, {Tissera}, {Tojeiro}, {Hernandez Toledo}, {de la Torre}, {Tremonti}, {Troup}, {Valenzuela}, {Martinez Valpuesta}, {Vargas-Gonz{\'a}lez}, {Vargas-Maga{\~n}a}, {Vazquez}, {Villanova}, {Vivek}, {Vogt},
  {Wake}, {Walterbos}, {Wang}, {Weaver}, {Weijmans}, {Weinberg}, {Westfall}, {Whelan}, {Wild}, {Wilson}, {Wood-Vasey}, {Wylezalek}, {Xiao}, {Yan}, {Yang}, {Ybarra}, {Y{\`e}che}, {Zakamska}, {Zamora}, {Zarrouk}, {Zasowski}, {Zhang}, {Zhao}, {Zheng}, {Zheng}, {Zhou}, {Zhou}, {Zhu}, {Zoccali}, \& {Zou}}]{2017AJ....154...28B}
{Blanton}, M.~R., {Bershady}, M.~A., {Abolfathi}, B., {et~al.} 2017, \aj, 154, 28, \dodoi{10.3847/1538-3881/aa7567}

\bibitem[{{Caon} {et~al.}(1993){Caon}, {Capaccioli}, \& {D'Onofrio}}]{caon}
{Caon}, N., {Capaccioli}, M., \& {D'Onofrio}, M. 1993, \mnras, 265, 1013, \dodoi{10.1093/mnras/265.4.1013}

\bibitem[{{Capuzzo-Dolcetta} \& {Mastrobuono-Battisti}(2009)}]{CDMB2009}
{Capuzzo-Dolcetta}, R., \& {Mastrobuono-Battisti}, A. 2009, \aap, 507, 183, \dodoi{10.1051/0004-6361/200912255}

\bibitem[{Carlsten {et~al.}(2022)Carlsten, Greene, Beaton, \& Greco}]{Carlsten_2022}
Carlsten, S.~G., Greene, J.~E., Beaton, R.~L., \& Greco, J.~P. 2022, The Astrophysical Journal, 927, 44, \dodoi{10.3847/1538-4357/ac457e}

\bibitem[{{C{\^o}t{\'e}} {et~al.}(2006){C{\^o}t{\'e}}, {Piatek}, {Ferrarese}, {Jord{\'a}n}, {Merritt}, {Peng}, {Ha{\c{s}}egan}, {Blakeslee}, {Mei}, {West}, {Milosavljevi{\'c}}, \& {Tonry}}]{Cote2006}
{C{\^o}t{\'e}}, P., {Piatek}, S., {Ferrarese}, L., {et~al.} 2006, \apjs, 165, 57, \dodoi{10.1086/504042}

\bibitem[{{Crnojevi{\'c}} {et~al.}(2016){Crnojevi{\'c}}, {Sand}, {Zaritsky}, {Spekkens}, {Willman}, \& {Hargis}}]{Crnojevic2016}
{Crnojevi{\'c}}, D., {Sand}, D.~J., {Zaritsky}, D., {et~al.} 2016, \apjl, 824, L14, \dodoi{10.3847/2041-8205/824/1/L14}

\bibitem[{{de Vaucouleurs}(1948)}]{dev}
{de Vaucouleurs}, G. 1948, Annales d'Astrophysique, 11, 247

\bibitem[{{den Brok} {et~al.}(2014){den Brok}, {Peletier}, {Seth}, {Balcells}, {Dominguez}, {Graham}, {Carter}, {Erwin}, {Ferguson}, {Goudfrooij}, {Guzm{\'a}n}, {Hoyos}, {Jogee}, {Lucey}, {Phillipps}, {Puzia}, {Valentijn}, {Verdoes Kleijn}, \& {Weinzirl}}]{denBrok2014}
{den Brok}, M., {Peletier}, R.~F., {Seth}, A., {et~al.} 2014, \mnras, 445, 2385, \dodoi{10.1093/mnras/stu1906}

\bibitem[{{Dey} {et~al.}(2019){Dey}, {Schlegel}, {Lang}, {Blum}, {Burleigh}, {Fan}, {Findlay}, {Finkbeiner}, {Herrera}, {Juneau}, {Landriau}, {Levi}, {McGreer}, {Meisner}, {Myers}, {Moustakas}, {Nugent}, {Patej}, {Schlafly}, {Walker}, {Valdes}, {Weaver}, {Y{\`e}che}, {Zou}, {Zhou}, {Abareshi}, {Abbott}, {Abolfathi}, {Aguilera}, {Alam}, {Allen}, {Alvarez}, {Annis}, {Ansarinejad}, {Aubert}, {Beechert}, {Bell}, {BenZvi}, {Beutler}, {Bielby}, {Bolton}, {Brice{\~n}o}, {Buckley-Geer}, {Butler}, {Calamida}, {Carlberg}, {Carter}, {Casas}, {Castander}, {Choi}, {Comparat}, {Cukanovaite}, {Delubac}, {DeVries}, {Dey}, {Dhungana}, {Dickinson}, {Ding}, {Donaldson}, {Duan}, {Duckworth}, {Eftekharzadeh}, {Eisenstein}, {Etourneau}, {Fagrelius}, {Farihi}, {Fitzpatrick}, {Font-Ribera}, {Fulmer}, {G{\"a}nsicke}, {Gaztanaga}, {George}, {Gerdes}, {Gontcho}, {Gorgoni}, {Green}, {Guy}, {Harmer}, {Hernand ez}, {Honscheid}, {Huang}, {James}, {Jannuzi}, {Jiang}, {Joyce}, {Karcher}, {Karkar}, {Kehoe}, {Kneib}, {Kueter-Young}, {Lan},
  {Lauer}, {Le Guillou}, {Le Van Suu}, {Lee}, {Lesser}, {Perreault Levasseur}, {Li}, {Mann}, {Marshall}, {Mart{\'\i}nez-V{\'a}zquez}, {Martini}, {du Mas des Bourboux}, {McManus}, {Meier}, {M{\'e}nard}, {Metcalfe}, {Mu{\~n}oz-Guti{\'e}rrez}, {Najita}, {Napier}, {Narayan}, {Newman}, {Nie}, {Nord}, {Norman}, {Olsen}, {Paat}, {Palanque-Delabrouille}, {Peng}, {Poppett}, {Poremba}, {Prakash}, {Rabinowitz}, {Raichoor}, {Rezaie}, {Robertson}, {Roe}, {Ross}, {Ross}, {Rudnick}, {Safonova}, {Saha}, {S{\'a}nchez}, {Savary}, {Schweiker}, {Scott}, {Seo}, {Shan}, {Silva}, {Slepian}, {Soto}, {Sprayberry}, {Staten}, {Stillman}, {Stupak}, {Summers}, {Sien Tie}, {Tirado}, {Vargas-Maga{\~n}a}, {Vivas}, {Wechsler}, {Williams}, {Yang}, {Yang}, {Yapici}, {Zaritsky}, {Zenteno}, {Zhang}, {Zhang}, {Zhou}, \& {Zhou}}]{dey}
{Dey}, A., {Schlegel}, D.~J., {Lang}, D., {et~al.} 2019, \aj, 157, 168, \dodoi{10.3847/1538-3881/ab089d}

\bibitem[{{Drinkwater} {et~al.}(2003){Drinkwater}, {Gregg}, {Hilker}, {Bekki}, {Couch}, {Ferguson}, {Jones}, \& {Phillipps}}]{Drinkwater2003}
{Drinkwater}, M.~J., {Gregg}, M.~D., {Hilker}, M., {et~al.} 2003, \nat, 423, 519, \dodoi{10.1038/nature01666}

\bibitem[{{Drinkwater} {et~al.}(2000){Drinkwater}, {Jones}, {Gregg}, \& {Phillipps}}]{Drinkwater2000}
{Drinkwater}, M.~J., {Jones}, J.~B., {Gregg}, M.~D., \& {Phillipps}, S. 2000, \pasa, 17, 227, \dodoi{10.1071/AS00034}

\bibitem[{{Dumont} {et~al.}(2022){Dumont}, {Seth}, {Strader}, {Voggel}, {Sand}, {Hughes}, {Caldwell}, {Crnojevi{\'c}}, {Mateo}, {Bailey}, \& {Forbes}}]{Dumont2022}
{Dumont}, A., {Seth}, A.~C., {Strader}, J., {et~al.} 2022, \apj, 929, 147, \dodoi{10.3847/1538-4357/ac551c}

\bibitem[{{El-Badry} {et~al.}(2023){El-Badry}, {Shen}, {Chandra}, {Bauer}, {Fuller}, {Strader}, {Chomiuk}, {Naidu}, {Caiazzo}, {Rodriguez}, {Nagarajan}, {Yamaguchi}, {Vanderbosch}, {Roulston}, {G{\"a}nsicke}, {Han}, {Burdge}, {Filippenko}, {Brink}, \& {Zheng}}]{2023OJAp....6E..28E}
{El-Badry}, K., {Shen}, K.~J., {Chandra}, V., {et~al.} 2023, The Open Journal of Astrophysics, 6, 28, \dodoi{10.21105/astro.2306.03914}

\bibitem[{{Forbes} {et~al.}(2018){Forbes}, {Read}, {Gieles}, \& {Collins}}]{forbes}
{Forbes}, D.~A., {Read}, J.~I., {Gieles}, M., \& {Collins}, M. L.~M. 2018, \mnras, 481, 5592, \dodoi{10.1093/mnras/sty2584}

\bibitem[{{Gadotti}(2009)}]{gadotti}
{Gadotti}, D.~A. 2009, \mnras, 393, 1531, \dodoi{10.1111/j.1365-2966.2008.14257.x}

\bibitem[{{Georgiev} \& {B{\"o}ker}(2014{\natexlab{a}})}]{Georgiev2014}
{Georgiev}, I.~Y., \& {B{\"o}ker}, T. 2014{\natexlab{a}}, \mnras, 441, 3570, \dodoi{10.1093/mnras/stu797}

\bibitem[{{Georgiev} \& {B{\"o}ker}(2014{\natexlab{b}})}]{georgiev}
---. 2014{\natexlab{b}}, \mnras, 441, 3570, \dodoi{10.1093/mnras/stu797}

\bibitem[{{Ginsburg} {et~al.}(2019){Ginsburg}, {Sip{\H{o}}cz}, {Brasseur}, {Cowperthwaite}, {Craig}, {Deil}, {Guillochon}, {Guzman}, {Liedtke}, {Lian Lim}, {Lockhart}, {Mommert}, {Morris}, {Norman}, {Parikh}, {Persson}, {Robitaille}, {Segovia}, {Singer}, {Tollerud}, {de Val-Borro}, {Valtchanov}, {Woillez}, {Astroquery Collaboration}, \& {a subset of astropy Collaboration}}]{astroquery}
{Ginsburg}, A., {Sip{\H{o}}cz}, B.~M., {Brasseur}, C.~E., {et~al.} 2019, \aj, 157, 98, \dodoi{10.3847/1538-3881/aafc33}

\bibitem[{{Gnedin} {et~al.}(2014){Gnedin}, {Ostriker}, \& {Tremaine}}]{gnedin}
{Gnedin}, O.~Y., {Ostriker}, J.~P., \& {Tremaine}, S. 2014, \apj, 785, 71, \dodoi{10.1088/0004-637X/785/1/71}

\bibitem[{{Green}(2018)}]{green}
{Green}, G. 2018, The Journal of Open Source Software, 3, 695, \dodoi{10.21105/joss.00695}

\bibitem[{Grossauer {et~al.}(2015)Grossauer, Taylor, Ferrarese, MacArthur, C√¥t√©, Roediger, Courteau, Cuillandre, Duc, Durrell, Gwyn, Jord√°n, Mei, \& Peng}]{grossauer}
Grossauer, J., Taylor, J.~E., Ferrarese, L., {et~al.} 2015, The Astrophysical Journal, 807, 88, \dodoi{10.1088/0004-637X/807/1/88}

\bibitem[{{Habas} {et~al.}(2020){Habas}, {Marleau}, {Duc}, {Durrell}, {Paudel}, {Poulain}, {S{\'a}nchez-Janssen}, {Sreejith}, {Ramasawmy}, {Stemock}, {Leach}, {Cuillandre}, {Gwyn}, {Agnello}, {B{\'\i}lek}, {Fensch}, {M{\"u}ller}, {Peng}, \& {van der Burg}}]{2020MNRAS.491.1901H}
{Habas}, R., {Marleau}, F.~R., {Duc}, P.-A., {et~al.} 2020, \mnras, 491, 1901, \dodoi{10.1093/mnras/stz3045}

\bibitem[{{Harris} {et~al.}(2017){Harris}, {Blakeslee}, \& {Harris}}]{Harris2017}
{Harris}, W.~E., {Blakeslee}, J.~P., \& {Harris}, G. L.~H. 2017, \apj, 836, 67, \dodoi{10.3847/1538-4357/836/1/67}

\bibitem[{{Hartmann} {et~al.}(2011){Hartmann}, {Debattista}, {Seth}, {Cappellari}, \& {Quinn}}]{Hartmann2011}
{Hartmann}, M., {Debattista}, V.~P., {Seth}, A., {Cappellari}, M., \& {Quinn}, T.~R. 2011, \mnras, 418, 2697, \dodoi{10.1111/j.1365-2966.2011.19659.x}

\bibitem[{{Hilker} {et~al.}(1999){Hilker}, {Infante}, \& {Richtler}}]{Hilker1999}
{Hilker}, M., {Infante}, L., \& {Richtler}, T. 1999, \aaps, 138, 55, \dodoi{10.1051/aas:1999495}

\bibitem[{{Hinshaw} {et~al.}(2013){Hinshaw}, {Larson}, {Komatsu}, {Spergel}, {Bennett}, {Dunkley}, {Nolta}, {Halpern}, {Hill}, \& {Odegard}}]{wmap9}
{Hinshaw}, G., {Larson}, D., {Komatsu}, E., {et~al.} 2013, \apjs, 208, 19, \dodoi{10.1088/0067-0049/208/2/19}

\bibitem[{{Hoyer} {et~al.}(2023){Hoyer}, {Neumayer}, {Seth}, {Georgiev}, \& {Greene}}]{hoyer23}
{Hoyer}, N., {Neumayer}, N., {Seth}, A.~C., {Georgiev}, I.~Y., \& {Greene}, J.~E. 2023, \mnras, 520, 4664, \dodoi{10.1093/mnras/stad220}

\bibitem[{{Hunter}(2007)}]{matplotlib}
{Hunter}, J.~D. 2007, Computing in Science and Engineering, 9, 90, \dodoi{10.1109/MCSE.2007.55}

\bibitem[{{Isobe} {et~al.}(1990){Isobe}, {Feigelson}, {Akritas}, \& {Babu}}]{isobe}
{Isobe}, T., {Feigelson}, E.~D., {Akritas}, M.~G., \& {Babu}, G.~J. 1990, \apj, 364, 104, \dodoi{10.1086/169390}

\bibitem[{{Jord{\'a}n} {et~al.}(2009){Jord{\'a}n}, {Peng}, {Blakeslee}, {C{\^o}t{\'e}}, {Eyheramendy}, {Ferrarese}, {Mei}, {Tonry}, \& {West}}]{2009Jordan_GC_NSC_Lum}
{Jord{\'a}n}, A., {Peng}, E.~W., {Blakeslee}, J.~P., {et~al.} 2009, \apjs, 180, 54, \dodoi{10.1088/0067-0049/180/1/54}

\bibitem[{{Kadowaki} {et~al.}(2021){Kadowaki}, {Zaritsky}, {Donnerstein}, {RS}, {Karunakaran}, \& {Spekkens}}]{kadowaki21}
{Kadowaki}, J., {Zaritsky}, D., {Donnerstein}, R.~L., {et~al.} 2021, \apj, 923, 257, \dodoi{10.3847/1538-4357/ac2948}

\bibitem[{{Lambert} {et~al.}(2024){Lambert}, {Khim}, {Zaritsky}, \& {Donnerstein}}]{lambert}
{Lambert}, M., {Khim}, D.~J., {Zaritsky}, D., \& {Donnerstein}, R. 2024, \aj, 167, 61, \dodoi{10.3847/1538-3881/ad0f25}

\bibitem[{{Lauer} {et~al.}(2005){Lauer}, {Faber}, {Gebhardt}, {Richstone}, {Tremaine}, {Ajhar}, {Aller}, {Bender}, {Dressler}, {Filippenko}, {Green}, {Grillmair}, {Ho}, {Kormendy}, {Magorrian}, {Pinkney}, \& {Siopis}}]{Lauer2005}
{Lauer}, T.~R., {Faber}, S.~M., {Gebhardt}, K., {et~al.} 2005, \aj, 129, 2138, \dodoi{10.1086/429565}

\bibitem[{{Li} {et~al.}(2023){Li}, {Greene}, {Greco}, {Huang}, {Melchior}, {Beaton}, {Casey}, {Danieli}, {Goulding}, {Joseph}, {Kado-Fong}, {Kim}, \& {MacArthur}}]{Li2023}
{Li}, J., {Greene}, J.~E., {Greco}, J.~P., {et~al.} 2023, \apj, 955, 1, \dodoi{10.3847/1538-4357/ace829}

\bibitem[{{Lim} {et~al.}(2018){Lim}, {Peng}, {C{\^o}t{\'e}}, {Sales}, {den Brok}, {Blakeslee}, \& {Guhathakurta}}]{Lim2018}
{Lim}, S., {Peng}, E.~W., {C{\^o}t{\'e}}, P., {et~al.} 2018, \apj, 862, 82, \dodoi{10.3847/1538-4357/aacb81}

\bibitem[{{Lim} {et~al.}(2020){Lim}, {C{\^o}t{\'e}}, {Peng}, {Ferrarese}, {Roediger}, {Durrell}, {Mihos}, {Wang}, {Gwyn}, {Cuillandre}, {Liu}, {S{\'a}nchez-Janssen}, {Toloba}, {Sales}, {Guhathakurta}, {Lan{\c{c}}on}, \& {Puzia}}]{lim}
{Lim}, S., {C{\^o}t{\'e}}, P., {Peng}, E.~W., {et~al.} 2020, \apj, 899, 69, \dodoi{10.3847/1538-4357/aba433}

\bibitem[{{Lisker} {et~al.}(2007){Lisker}, {Grebel}, {Binggeli}, \& {Glatt}}]{Lisker07}
{Lisker}, T., {Grebel}, E.~K., {Binggeli}, B., \& {Glatt}, K. 2007, \apj, 660, 1186, \dodoi{10.1086/513090}

\bibitem[{{Lotz} {et~al.}(2001){Lotz}, {Telford}, {Ferguson}, {Miller}, {Stiavelli}, \& {Mack}}]{lotz}
{Lotz}, J.~M., {Telford}, R., {Ferguson}, H.~C., {et~al.} 2001, \apj, 552, 572, \dodoi{10.1086/320545}

\bibitem[{{McKinney}(2010)}]{pandas}
{McKinney}, W. 2010, Proceedings of the 9th Python in Science Conference, 51

\bibitem[{{Mihos} \& {Hernquist}(1994)}]{mihos94}
{Mihos}, J.~C., \& {Hernquist}, L. 1994, \apjl, 437, L47, \dodoi{10.1086/187679}

\bibitem[{{Millman} \& {Aivazis}(2011)}]{scipy2}
{Millman}, K.~J., \& {Aivazis}, M. 2011, Computing in Science and Engineering, 13, 9, \dodoi{10.1109/MCSE.2011.36}

\bibitem[{{Moustakas} {et~al.}(2023){Moustakas}, {Lang}, {Dey}, {Juneau}, {Meisner}, {Myers}, {Schlafly}, {Schlegel}, {Valdes}, {Weaver}, \& {Zhou}}]{moustakas}
{Moustakas}, J., {Lang}, D., {Dey}, A., {et~al.} 2023, \apjs, 269, 3, \dodoi{10.3847/1538-4365/acfaa2}

\bibitem[{{Neumayer} {et~al.}(2020){Neumayer}, {Seth}, \& {B{\"o}ker}}]{neumayer}
{Neumayer}, N., {Seth}, A., \& {B{\"o}ker}, T. 2020, \aapr, 28, 4, \dodoi{10.1007/s00159-020-00125-0}

\bibitem[{{Neumayer} \& {Walcher}(2012)}]{Neumayer2012}
{Neumayer}, N., \& {Walcher}, C.~J. 2012, Advances in Astronomy, 2012, 709038, \dodoi{10.1155/2012/709038}

\bibitem[{{Norris} {et~al.}(2014){Norris}, {Kannappan}, {Forbes}, {Romanowsky}, {Brodie}, {Faifer}, {Huxor}, {Maraston}, {Moffett}, {Penny}, {Pota}, {Smith-Castelli}, {Strader}, {Bradley}, {Eckert}, {Fohring}, {McBride}, {Stark}, \& {Vaduvescu}}]{Norris2014}
{Norris}, M.~A., {Kannappan}, S.~J., {Forbes}, D.~A., {et~al.} 2014, \mnras, 443, 1151, \dodoi{10.1093/mnras/stu1186}

\bibitem[{{Oke}(1964)}]{oke1}
{Oke}, J.~B. 1964, \apj, 140, 689, \dodoi{10.1086/147960}

\bibitem[{{Oke} \& {Gunn}(1983)}]{oke2}
{Oke}, J.~B., \& {Gunn}, J.~E. 1983, \apj, 266, 713, \dodoi{10.1086/160817}

\bibitem[{{Oliphant}(2007)}]{scipy1}
{Oliphant}, T.~E. 2007, Computing in Science and Engineering, 9, 10, \dodoi{10.1109/MCSE.2007.58}

\bibitem[{{Ordenes-Brice{\~n}o} {et~al.}(2018){Ordenes-Brice{\~n}o}, {Puzia}, {Eigenthaler}, {Taylor}, {Mu{\~n}oz}, {Zhang}, {Alamo-Mart{\'\i}nez}, {Ribbeck}, {Grebel}, {{\'A}ngel}, {C{\^o}t{\'e}}, {Ferrarese}, {Hilker}, {Lan{\c{c}}on}, {Mieske}, {Miller}, {Rong}, \& {S{\'a}nchez-Janssen}}]{Ordenes2018}
{Ordenes-Brice{\~n}o}, Y., {Puzia}, T.~H., {Eigenthaler}, P., {et~al.} 2018, \apj, 860, 4, \dodoi{10.3847/1538-4357/aac1b8}

\bibitem[{{Peng} {et~al.}(2002){Peng}, {Ho}, {Impey}, \& {Rix}}]{peng}
{Peng}, C.~Y., {Ho}, L.~C., {Impey}, C.~D., \& {Rix}, H.-W. 2002, \aj, 124, 266, \dodoi{10.1086/340952}

\bibitem[{Peng {et~al.}(2010)Peng, Ho, Impey, \& Rix}]{Peng_2010}
Peng, C.~Y., Ho, L.~C., Impey, C.~D., \& Rix, H.-W. 2010, The Astronomical Journal, 139, 2097, \dodoi{10.1088/0004-6256/139/6/2097}

\bibitem[{{Pfeffer} \& {Baumgardt}(2013)}]{Pfeffer2013}
{Pfeffer}, J., \& {Baumgardt}, H. 2013, \mnras, 433, 1997, \dodoi{10.1093/mnras/stt867}

\bibitem[{{Roediger} \& {Courteau}(2015)}]{roediger}
{Roediger}, J.~C., \& {Courteau}, S. 2015, \mnras, 452, 3209, \dodoi{10.1093/mnras/stv1499}

\bibitem[{{S{\'a}nchez-Janssen} {et~al.}(2019){S{\'a}nchez-Janssen}, {Puzia}, {Ferrarese}, {C{\^o}t{\'e}}, {Eigenthaler}, {Miller}, {Ordenes-Brice{\~n}o}, {Peng}, {Ribbeck}, {Roediger}, {Spengler}, \& {Taylor}}]{sanchez2019b}
{S{\'a}nchez-Janssen}, R., {Puzia}, T.~H., {Ferrarese}, L., {et~al.} 2019, \mnras, 486, L1, \dodoi{10.1093/mnrasl/slz008}

\bibitem[{{Schlegel} {et~al.}(1998){Schlegel}, {Finkbeiner}, \& {Davis}}]{SFD}
{Schlegel}, D.~J., {Finkbeiner}, D.~P., \& {Davis}, M. 1998, \apj, 500, 525, \dodoi{10.1086/305772}

\bibitem[{{S{\'e}rsic}(1963)}]{sersic}
{S{\'e}rsic}, J.~L. 1963, Boletin de la Asociacion Argentina de Astronomia La Plata Argentina, 6, 41

\bibitem[{{Seth} {et~al.}(2006){Seth}, {Dalcanton}, {Hodge}, \& {Debattista}}]{Seth2006}
{Seth}, A.~C., {Dalcanton}, J.~J., {Hodge}, P.~W., \& {Debattista}, V.~P. 2006, \aj, 132, 2539, \dodoi{10.1086/508994}

\bibitem[{{Spitler} \& {Forbes}(2009)}]{Spitler2009}
{Spitler}, L.~R., \& {Forbes}, D.~A. 2009, \mnras, 392, L1, \dodoi{10.1111/j.1745-3933.2008.00567.x}

\bibitem[{Sugiura(1978)}]{sugiura}
Sugiura, N. 1978, Communications in Statistics - Theory and Methods, 7, 13, \dodoi{10.1080/03610927808827599}

\bibitem[{Sánchez-Janssen {et~al.}(2019)Sánchez-Janssen, Côté, Ferrarese, Peng, Roediger, Blakeslee, Emsellem, Puzia, Spengler, Taylor, Álamo Martínez, Boselli, Cantiello, Cuillandre, Duc, Durrell, Gwyn, MacArthur, Lançon, Lim, Liu, Mei, Miller, Muñoz, Mihos, Paudel, Powalka, \& Toloba}]{sanchez2019}
Sánchez-Janssen, R., Côté, P., Ferrarese, L., {et~al.} 2019, The Astrophysical Journal, 878, 18, \dodoi{10.3847/1538-4357/aaf4fd}

\bibitem[{{Tremaine} {et~al.}(1975){Tremaine}, {Ostriker}, \& {Spitzer}}]{tremaine1975}
{Tremaine}, S.~D., {Ostriker}, J.~P., \& {Spitzer}, L., J. 1975, \apj, 196, 407, \dodoi{10.1086/153422}

\bibitem[{{van der Walt} {et~al.}(2011){van der Walt}, {Colbert}, \& {Varoquaux}}]{numpy}
{van der Walt}, S., {Colbert}, S.~C., \& {Varoquaux}, G. 2011, Computing in Science and Engineering, 13, 22, \dodoi{10.1109/MCSE.2011.37}

\bibitem[{{van Dokkum} {et~al.}(2015{\natexlab{a}}){van Dokkum}, {Abraham}, {Merritt}, {Zhang}, {Geha}, \& {Conroy}}]{2015vanDokkum}
{van Dokkum}, P.~G., {Abraham}, R., {Merritt}, A., {et~al.} 2015{\natexlab{a}}, \apjl, 798, L45, \dodoi{10.1088/2041-8205/798/2/L45}

\bibitem[{{van Dokkum} {et~al.}(2015{\natexlab{b}}){van Dokkum}, {Abraham}, {Merritt}, {Zhang}, {Geha}, \& {Conroy}}]{vdk15a}
---. 2015{\natexlab{b}}, \apjl, 798, L45, \dodoi{10.1088/2041-8205/798/2/L45}

\bibitem[{{Walcher} {et~al.}(2006){Walcher}, {B{\"o}ker}, {Charlot}, {Ho}, {Rix}, {Rossa}, {Shields}, \& {van der Marel}}]{Walcher2006}
{Walcher}, C.~J., {B{\"o}ker}, T., {Charlot}, S., {et~al.} 2006, \apj, 649, 692, \dodoi{10.1086/505166}

\bibitem[{{Wang} {et~al.}(2023){Wang}, {Peng}, {Liu}, {Mihos}, {C{\^o}t{\'e}}, {Ferrarese}, {Taylor}, {Blakeslee}, {Cuillandre}, {Duc}, {Guhathakurta}, {Gwyn}, {Ko}, {Lan{\c{c}}on}, {Lim}, {MacArthur}, {Puzia}, {Roediger}, {Sales}, {S{\'a}nchez-Janssen}, {Spengler}, {Toloba}, {Zhang}, \& {Zhu}}]{Wang+23}
{Wang}, K., {Peng}, E.~W., {Liu}, C., {et~al.} 2023, \nat, 623, 296, \dodoi{10.1038/s41586-023-06650-z}

\bibitem[{{Zaritsky}(2022)}]{2022Zaritsky_GC}
{Zaritsky}, D. 2022, \mnras, 513, 2609, \dodoi{10.1093/mnras/stac1072}

\bibitem[{{Zaritsky} \& {Behroozi}(2023)}]{2023MNRAS.519..871Z}
{Zaritsky}, D., \& {Behroozi}, P. 2023, \mnras, 519, 871, \dodoi{10.1093/mnras/stac3610}

\bibitem[{{Zaritsky} {et~al.}(2016){Zaritsky}, {Crnojevi{\'c}}, \& {Sand}}]{zaritsky16}
{Zaritsky}, D., {Crnojevi{\'c}}, D., \& {Sand}, D.~J. 2016, \apjl, 826, L9, \dodoi{10.3847/2041-8205/826/1/L9}

\bibitem[{{Zaritsky} {et~al.}(2023){Zaritsky}, {Donnerstein}, {Dey}, {Karunakaran}, {Kadowaki}, {Khim}, {Spekkens}, \& {Zhang}}]{smudges5}
{Zaritsky}, D., {Donnerstein}, R., {Dey}, A., {et~al.} 2023, \apjs, 267, 27, \dodoi{10.3847/1538-4365/acdd71}

\bibitem[{{Zaritsky} {et~al.}(2021){Zaritsky}, {Donnerstein}, {Karunakaran}, {Barbosa}, {Dey}, {Kadowaki}, {Spekkens}, \& {Zhang}}]{smudges2}
{Zaritsky}, D., {Donnerstein}, R., {Karunakaran}, A., {et~al.} 2021, \apjs, 257, 60, \dodoi{10.3847/1538-4365/ac2607}

\bibitem[{{Zaritsky} {et~al.}(2022){Zaritsky}, {Donnerstein}, {Karunakaran}, {Barbosa}, {Dey}, {Kadowaki}, {Spekkens}, \& {Zhang}}]{smudges3}
---. 2022, \apjs, 261, 11, \dodoi{10.3847/1538-4365/ac6ceb}

\bibitem[{{Zaritsky} {et~al.}(2019){Zaritsky}, {Donnerstein}, {Dey}, {Kadowaki}, {Zhang}, {Karunakaran}, {Mart{\'\i}nez-Delgado}, {Rahman}, \& {Spekkens}}]{smudges}
{Zaritsky}, D., {Donnerstein}, R., {Dey}, A., {et~al.} 2019, \apjs, 240, 1, \dodoi{10.3847/1538-4365/aaefe9}

\end{thebibliography}
\bibliographystyle{aasjournal}

\end{document}